\documentclass[oldversion]{aa}
\usepackage{graphicx}
\usepackage{epsfig}
\usepackage{color}
\usepackage{natbib}
\bibpunct{(}{)}{;}{a}{}{,} 

\begin{document}

\title{The dilution peak, metallicity evolution, and dating of galaxy interactions and mergers}

\titlerunning{Metallicity Evolution in Galaxy Interactions and Mergers}

\author{Marco Montuori$^{1}$, Paola Di Matteo$^{2}$, Matthew D. Lehnert$^{2}$,
 Fran\c coise Combes$^{3}$, Benoit Semelin$^{3}$}

\authorrunning{Montuori et al.}

\institute{$^{1}$ SMC-ISC-CNR \& Physics Department, Univ. Roma "La Sapienza", Pl. Aldo Moro 2, 00185, Rome, Italy\\ 
$^{2}$  Observatoire de Paris, section de Meudon, GEPI,
5 Place Jules Jannsen, 92195, Meudon, France\\
$^{3}$ Observatoire de Paris, LERMA, CNRS, UPMC, 61 Avenue de l'Observatoire, 75014 Paris, France}

\date{Accepted, Received}

\abstract{Strong inflows of gas from the outer disk to the inner
kiloparsecs are induced during the interaction of disk galaxies. This 
inflow of relatively low-metallicity gas dilutes the metallicity
of the circumnuclear gas. This process is critical for the galaxy evolutions.
We have investigated several aspects of the process as the timing and 
duration of the dilution and its correlation with the induced star formation. 
We analysed major (1:1) gas-rich interactions and mergers,
spanning a range of initial orbital characteristics. Star formation 
and metal enrichment from SNe are included
in our model. Our results show that the strongest trend is between
the star formation rate and the dilution of the metals in the nuclear
region; i.e., the more intense the central burst of star formation,
the more the gas is diluted. This trend comes from strong inflows of
relatively metal-poor gas from the outer regions of both disks, which
fuels the intense star formation and lowers the overall metallicity
for a time.  The strong inflows happen on timescales of about 10$^8$
years or less (i.e., on an internal  dynamical time of the disk in the
simulations), and the most intense star formation and lowest gas phase
metallicities are seen generally after the first pericentre passage.
As the star formation proceeds and the merger advances, the dilution
reduces and enrichment becomes dominant -- ultimately increasing the
metallicity of the circumnuclear gas to a level higher than the initial
metallicities of the merging galaxies. The ``fly-bys'' --
pairs that interact but do not merge -- also cause some dilution.
We even see some dilution early in the merger or in the ``fly-bys'' and
thus do not observe a strong trend between the nuclear metallicities and
separation in our simulations until the merger is well advanced.  We also
analyse the O and Fe enrichment of the ISM, and show that the evolution
of the $\alpha/Fe$ ratios, as well as the dilution of the central gas
metallicity, can be used as a clock for ``dating'' the interaction.}


\maketitle

\section{Introduction}

Understanding how galaxy interactions and mergers shape the ensemble
population of galaxies in the local universe is crucial for our
understanding of galaxy evolution. It has been recognised for several
decades that mergers may play an important role in the evolution of
galaxies, especially for the early types \citep[e.g.,][]{toomre77}.
While there have been many observations and theoretical studies of
``first-order'' effects in interactions and mergers 
(such as morphological and kinetical evolutiona and star formation), much less work
has been done on 2$^{nd}$ order effects such as the evolution of
metallicity.

Recently, we have undertaken the study of the evolution of metallicity
in major dry (i.e., gas poor) mergers, showing that they can lead to
metallicity gradients in agreement with those of local ellipticals
\citep{dimatteo09}. However, in gas-poor mergers, the evolution of the
local galaxy metallicity is driven only by the mixing of stars during
the coalescence process. Gas can alter the properties of the final
remnant significantly, affecting both the final gas phase and stellar
metal abundance and ratios.  In particular, gas phase metallicity
can play an important role in constraining our understanding of the
complex interaction between gas flows and star formation in interactions
and mergers and allow us to constrain the relative time scales of the different epochs (i.e.,
date the merger/interaction).

If the progenitor disks have strong gas metallicity gradients, as
observed in many galaxies in the local universe \citep{shields90,
diner96}, one would expect that interaction-induced gas inflows will
drive a noticeable amount of gas into the central regions from
the outer disk and thus initially lower the circumnuclear gas phase
metallicity. Outflows of gas ejected in tidal tails may contribute
to modifying the metallicity profile of the final merger remnant.
Subsequent star formation (and outflows) may increase (decrease) the
gas phase circumnuclear metallicity.  The exact timing of the inflows
versus the initiation of intense star formation and subsequent metal
enrichment may provide robust constraints on the underlying physical
mechanisms that determine the rate and relative timing of gas flows
and star formation within the models.  Comparing the evolution of the
metallicity and its relationship to both inflows and star formation will
allow us to constrain the dissipation time scales, the star formation evolution, 
the time delay between the onset of inflows,
sufficient to dilute the metallicity, etc.

In the past few years, a number of studies have investigated the role
played by interactions and environment in determining the gas phase
metallicities of galaxies. \citet{donzelli00} show that merging
galaxies have on average higher excitation in their optical emission
line spectra than interacting pairs, and attribute this difference
to lower gas metallicity in the mergers.  \citet{marquez02} studied a
sample of more than one hundred spiral galaxies, ranging from isolated
to interaction with strong morphological distortions, and show
that the [NII]/H$\alpha$ ratios, used as a metallicity indicator, indicate 
a clear trend from the metallicity to morphological type, with
earlier type spirals showing higher ratios, while they found no trend
with the status of the interaction. Recently, \citet{kewley06} have derived
the luminosity-metallicity relation for a sample of local galaxy pairs and
compared it with that of nearby field galaxies. They found that pairs with
small projected separations (s $< 20\; \mathrm{kpc \; h^{-1}}$) have systematically
lower metallicities than either isolated galaxies or pairs with larger
separations, for a given luminosity. They also found a correlation
between gas metallicity and burst strengths -- all  galaxies in their
interacting sample with strong central bursts having close companions
and metallicities lower than the comparable field galaxies or pairs with
wider separations.  \citet{rupke08}, in a study of strong interactions with
high star formation rates (ultra-luminous infrared galaxies), found that
the metal abundance in these intense starbursts is a factor of two lower
than that of galaxies of comparable luminosity and mass. These results
have been generally explained by gas inflows induced by the interaction,
which  dilute the pre-existing nuclear gas to produce a lower metallicity
than that observed for wider separated pairs or isolated systems.
More recently, \citet{peeples08, peeples09} have studied, respectively,
a sample of low-mass, high-metallicity and high-mass, low-metallicity
outliers from the mass-metallicity relation of star-forming galaxies
from selected from the Sloan Digital Sky Survey (SDSS).They showed that
the low-mass, high-metallicity outliers are usually isolated galaxies,
with no evident companion or strong interactions. On the other hand, the
high-mass, low-metallicity outlierts typically consist of systems that
have high star formation rates and evidence for disturbed morphologies.

However, \citet{cooper08} found a strong metallicity-density relation for
star forming galaxies in the local universe, with the more metal-rich
galaxies apparently favouring regions of higher galaxy overdensity
\citep[see also][]{ellison09}. They conclude that the discrepancy found
with the outer studies (including those just discussed) is due to the
fact that the number of close pairs (s $< 100\; \mathrm{kpc\;h^{-1}}$) in the SDSS
sample constitutes only a tiny fraction of the whole sample (less than 1\%
of galaxies). In this case, close pairs with low central metallicity could
not contribute significantly to the scatter found in the mass-metallicity
relation.  Of course it can be difficult to discern trends in the gas
phase metallicity during a merger when studying large samples of galaxies
in SDSS as other factors may dominate the overall mass-metallicity
relationship and its scatter.  A confirmation that the mass-metallicity
relation is affected by interactions only for close pairs showing signs of
strong disturbances has been found by \citet{dansac08}, who pointed
out that, in such pairs, the gas metallicity depends on the mass ratio
of the two interacting systems: less massive members are systematically
enriched, while a galaxy in interaction with a comparable stellar mass
companion shows a metallicity lower than that of a galaxy in isolation.
So while mergers and interactions alone may not drive the overall scatter
in the mass-metallicity relationship, it may be a contributing factor.
All these studies suggest that the dilution and enrichment of the ISM in
the central regions of disk galaxies strongly depend on the exact timing
of the different processes at play: gas inflows, interaction-driven star
formation, gas consumption, feedback and subsequent enrichment.

While a number of numerical studies have investigated the response
of the gaseous component of galaxies during tidal interactions and
the subsequent star formation such interactions induce \citep{iono04,
mihos94a, springel00, cox06, cox08,  kapferer05, dimatteo07, dimatteo08},
little attention has been given to the detailed evolution of the metal
content during galaxy encounters. Using a galaxy pair catalogue from
cosmological simulations, \citet{perez06} have shown that the O/H
abundance ratio in the  central regions of close galaxy pairs 
(s $< 50\; \mathrm{kpc\;h^{-1}}$) shows a lower level of enrichment than the mean O/H
abundance ratio of a control sample, thus confirming the role played
by gas inflow in diluting the metal content of the nuclear regions.
Recently, \citet{rupke10} have analysed simulations of major galaxy
mergers, studying the dilution of the gas metallicity in the nuclear
regions due to gas inflow. They found a dilution of about 0.1 - 0.3 dex,
happening shortly after the first pericentre passage between the two
galaxies. However, their models do not include either star formation
prescriptions or metal enrichment, so that while it has been possible to
give predictions about the strength of the dilution, nothing is still
known about the role played by interaction-induced star formation in
the metal dilution and then subsequent enrichement.  Moreover, the exact
timing of the dilution peak and its correlation with the increase in the
amplitude of the star formation have not been studied yet in any detail.

\section{Models and initial conditions}

Here we study mergers and flybys involving two massive Sbc galaxies,
having a mass ratio 1:1. The Sbc galaxies (hereafter called gSb)
are composed of a spherical dark matter halo and a spherical
bulge, represented by Plummer spheres \citep{bt1} with total masses
$M_H=1.7\times10^{11}\;\mathrm{M_{\sun}}$ and $M_B=11.5\times10^9\;\mathrm{M_{\sun}}$ and
core radii $r_H=12\;\mathrm{kpc}$ and $r_B=1\;\mathrm{kpc}$ respectively. The stellar disk is
represented by a Miyamoto-Nagai density profile \citep{bt1} with mass
$M_{*}=40.6\times10^9\;\mathrm{M_{\sun}}$, vertical and radial scale lengths
given respectively by $h_{*}=0.5\;\mathrm{kpc}$ and $a_{*}=5\;\mathrm{kpc}$. The galaxies
initially contain a gas mass  $M_{gas}=0.2 \; M_{*}$, redistributed in
a Miyamoto-Nagai disk with vertical and radial scale lengths given
respectively by $h_{gas}=0.2\;\mathrm{kpc}$ and $a_{gas}=6\;\mathrm{kpc}$. 
 For each pair of interacting galaxies, we performed
24  simulations, varying the galaxies orbital initial conditions
(initial orbital energy E and angular momentum L) and taking into account
both direct and retrograde orbits.
The initial orbital parameters (initial distance between the two galaxies, initial relative velocity, specific angular momentum, orbital energy and spin) for the different runs are fully described in Table 7 of \citet{chili10}, and we refer the reader to this paper for their complete description.

We chose a reference frame with its origin at the barycentre of the
system and x-y plane corresponding to the orbital plane.  For each
interacting pair, we have kept the disk of one of the two galaxies in
the orbital plane ($i_1 = 0 \degr$), and varied the inclination $i_2$ of
the companion disk, considering: $i_2 = 0 \degr$, $i_2 = 45 \degr$, $i_2 =
75 \degr$, and $i_2 = 90\degr$. The orbital angular momentum can be parallel
(direct orbit) or anti-parallel (retrograde orbit) to the z-axis of the
reference frame.

All the simulations (96 in total) were run using the Tree-SPH code
described in \citet{benoit02}. Each galaxy contains a total of
$N_{TOT}=120000$ particles, equally redistributed among gas, stars
and dark matter. A Plummer potential is used to soften gravity at
small scales, with constant softening lengths of $\epsilon=280$ pc for
all particles. The gas is modelled as isothermal, with a temperature
$T_{gas}= 10^4 \;\mathrm{K}$. The assumption of an isothermal
ISM is justified by the short cooling times in the gaseous disk at the
resolution of our models, such that fluctuations in the gas temperature,
those due to SN explosions for example, would be quicky radiated away
\citep[see the discussion in, e.g., ][]{mihos96}.

We refer the reader to \citet{benoit02} for a detail
description of and the tests performed to validate the code. Other tests
have been presented in  \citet{dimatteo08}. The equations of motion are
integrated using a leapfrog algorithm with a fixed time step of $\Delta
t = 5 \times 10^5\;\mathrm{yr}$.

\begin{figure*}
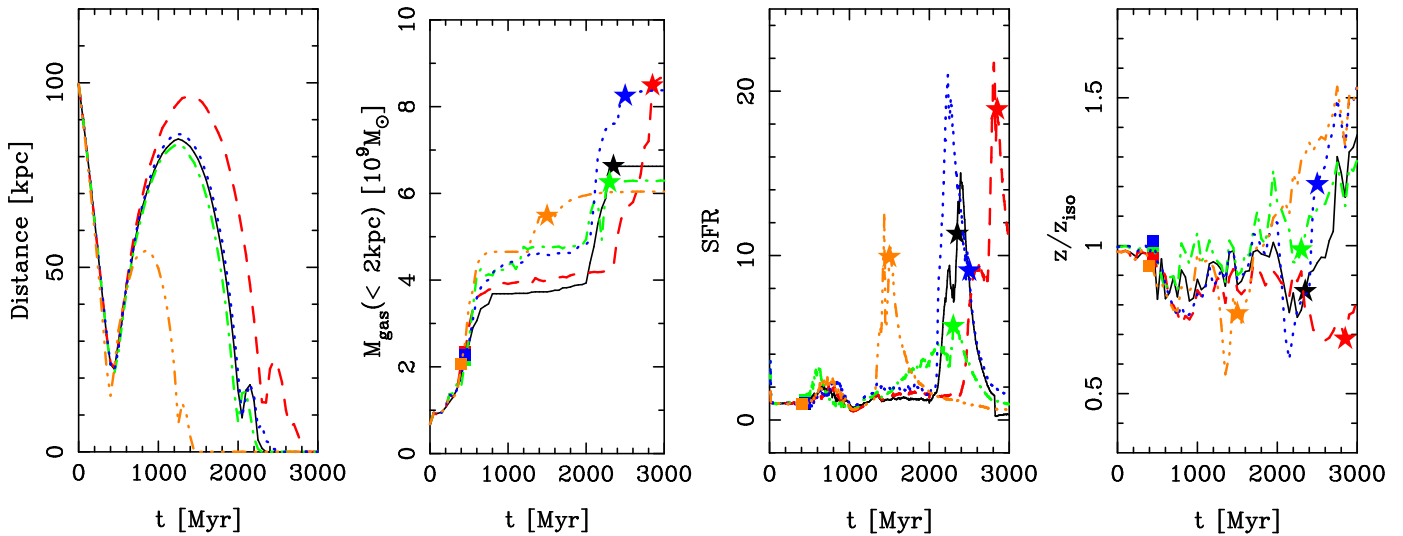

\centering
\includegraphics[width=7cm,angle=270]{14304fg1.ps}
\includegraphics[width=7cm,angle=270]{14304fg2.ps}
\includegraphics[width=7cm,angle=270]{14304fg3.ps}
\includegraphics[width=7cm,angle=270]{14304fg4.ps}
\caption{The dynamics of the gas during the interaction and subsequent
merger of two gSb galaxies for several orbits (shown with different
colours). {\it (First panel)}: Relative distance of the two galaxy centres
versus time for some simulated orbits. {\it (Second panel)}: Gas mass
inside 2 kpc from one of the two galaxy centres versus time. {\it (Third
panel)}:  Evolution of the star formation rate (SFR) relative to the
rate for the corresponding initial galaxies evolved in isolation. {\it
(Fourth panel)}: Evolution of the central gas metallicity. The metallicity
has been normalised to that of the initial galaxies used in the merger
simulations evolved isolated. In the second, third, and fourth panel,
squares and asterisks respresent respectively the time of the first
pericentre passage and the time of final coalescence for the different
orbits.}

\label{fig:timeevol}
\end{figure*}

\begin{figure*}
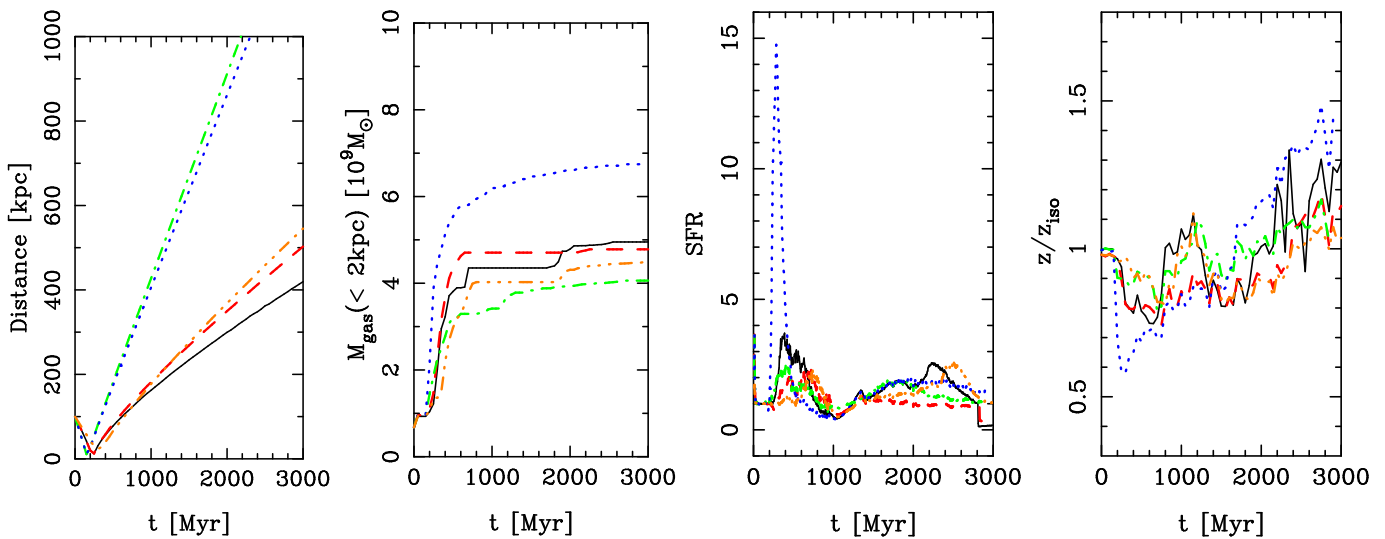

\centering
\includegraphics[width=7cm,angle=270]{14304fg5.ps}
\includegraphics[width=7cm,angle=270]{14304fg6.ps}
\includegraphics[width=7cm,angle=270]{14304fg7.ps}
\includegraphics[width=7cm,angle=270]{14304fg8.ps}
\caption{Same as Fig.\ref{fig:timeevol}, but for the simulations
of ``flybys''. Note that simulations resulting in starbursts
with higher intensities also have gas phase metallicities with higher
dilution.  Even in flybys, the star formation rate and dilution of the
metallicity are related.}
\label{fig:timeevol_fly}
\end{figure*}

\subsection{Star formation and metal enrichment}

The numerical recipes used to implement star formation and metal
enrichment have been fully described elsewhere  \citep{dimatteo07,
chili10}, thus here we only summarise those features that are most
directly relevant to our analysis. Star formation efficiency has been
parametrised as:

\begin{equation}\label{loc}
\frac{\dot{M}_{gas}}{M_{gas}}=C\times {\rho_{gas}}^{1/2}
\end{equation} 

where $M_{gas}$ is the gas mass of a particle, and $\rho_{gas}$ the local
gas density. The constant $C=0.3 \mathrm{ \; pc^{3/2} \; M_{\sun}^{-1/2}\;  Gyr^{-1}}$
has been chosen such that the isolated  disk galaxies form stars at
an average rate of $\sim 1 \;\mathrm{M_{\sun}\; yr^{-1}}$. Once the star formation
recipe is defined, we apply it to gas particles, using the hybrid method
described in \citet{mihos94b}.

The effect of star formation on the surrounding interstellar medium has
been implemented as follows.  For each star-forming hybrid particle,
we evaluate the fraction of stars formed with masses $> 8 \;\mathrm{M_{\sun}}$,
adopting a \citet{miller79} initial mass function (IMF), and we
assume that stars above this mass threshold instantaneously become
supernovae. Each of the SN leave behind a remnant of $1.4 \;\mathrm{M_{\sun}}$ and
releases their remaining mass to the surrounding ISM. The mass released
also enriches the surrounding gas with metals. This is done assuming
a yield $y=M_{ret}/M_{*}\;=\;0.02$, where $M_{ret}$ is the total mass of
all reprocessed metals and $M_{*}$ the total mass in stars. For each gas
particle, the return of mass and metals is applied to the $i-th$ neighbour
gas particle, using a weight $w_i$ based on the smoothing kernel. The
metallicity is initially distributed in a gradient of the form:

\begin{equation}\label{gradient}
z_{m}(R)= z_{0} \times 10^{-0.07 \,R}
\end{equation}

with $R$ the particle distance from the galaxy centre and $z_{0}=3 \,
z_{\sun}$ \citep{kennicutt03,magrini07,lemasle08}, evolves with time,
as star formation proceeds.

Supernovae explosions also inject energy into the
surrounding ISM. This is taken into account assuming that a fraction,
$\epsilon_{kin}$, of the energy, $E_{SN}$, released by a SNe goes into
kinetic energy giving a radial kick to the velocities of neighbouring
gas particles. The value of $\epsilon_{kin}$ has been chosen such that
the total amount of kinetic energy received by a gas particle due to
the contribution from all the surrounding neighbours, is less than 1 km
s$^{-1}$, thus preventing a rapid growth of the vertical thickness of
the gaseous disk. A more detailed description of the implementation of
feedback recipes used in our code is given in \citet{benoit02, chili10}.

\section{Results}

\subsection{Gas Flows: Where and When}

It is known that close passages of galaxies can destabilise galaxy disks,
producing asymmetries, such as bars, which are efficient mechanisms for
driving gas into the circumnuclear region of a galaxy, producing a burst
of star formation \citep{iono04, combes08}.  If, prior to the interaction,
the gas is distributed in the disk with a negative metallicity gradient,
with the outer regions being relatively more metal poor than the inner
ones as observed in  local galaxies \citep{shields90, diner96}, then a
portion of this low-metallicity gas will fall into the nuclear regions,
diluting the pre-existing gas with gas of lower metallicity. This
dilution will last until the star formation, which is enhanced by the
gas inflow, releases reprocessed metals to enrich the gas.  

This cycle (dilution and enrichment) is shown in Figs. \ref{fig:timeevol}
and \ref{fig:timeevol_fly} for some of our galaxy merger simulations.
For the analysis of the effect of mergers on the metallicity of
galaxies, we have defined the metallicity dilution (often referred to
as simply dilution) as $z/z_{iso}< 1$, where $z$ is the metallicity of
a galaxy in the pair during the merger or flyby and $z_{iso}$ is the
corresponding value for the same galaxy evolved isolated.  Both $z$ and
$z_{iso}$ are measured in apertures of 1-2 kpc and thus represent the
evolution of the metallicity for the nuclear or circumnuclear regions
of galaxies (which also corresponds reasonably well with the aperture
sizes used to estimate the metallicities of galaxies; see discussion in
the \S 1).

We find that as soon as the two galaxies have undergone their first
pericentre passage, an intense inflow of gas, lasting for 300-400
Myr, takes place. A detailed discussion of the mechanism
generating the gas inflows we observe in our simulations is given in
\citet{dimatteo07}.  They show that each close passage
of the two interacting galaxies is accompanied by the amplification
of stellar asymmetries -- stellar bars.  These stellar bars 
are responsible of removing
angular momentum from gas, transferring it to the stellar component, and
in regulating the subsequent inflow of gas in the inner kpc.  We note that
this result has already been discussed in several previous studies using
simulations (see
for example Mihos \& Hernquist 1994, 1996).  This mechanism does not require any stellar
feedback to work, that is, the gas inflows are a natural consequence of the
dynamics of the interaction and the redistribution of the angular momentum that
is induced during the progression of the merger
\citep{dimatteo08, rupke10}. This intense inflows can deposit $in few
10^8 ys$ into the central ($r\le 2\;\mathrm{kpc}$) regions $\sim 2-3
\times 10^9 \;\mathrm{M_{\sun}}$ of gas,  originally located before
the interaction at an average distance of $6-7\;\mathrm{kpc}$ from the
galaxy centre. The amount and origin of the inflowing gas obviously
depends on characteristics of the merger like orbital parameters of the
interaction and gas fraction \citep[see][for a discussion]{dimatteo07}.
In all cases, this inflow of gas fuels the first increase in the star
formation rate during the merging process.  A second, sometimes even
stronger peak of SF takes place in the final phases of the interaction
when the two galaxies are close to coalescence.

\begin{figure*}
\centering
\includegraphics[width=6cm,angle=-90]{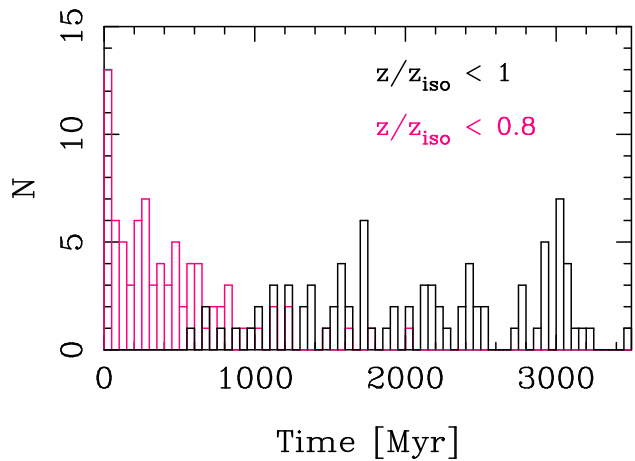}
\vspace{0.8cm}
\caption{The duration that the circumnuclear metallicity
is below 0.8 (shown in red) and between 0.8 and 1.0 (shown in black)
relative to an isolated simulated spiral with the same properties as
the individual merging galaxies.  We chose 0.8 as the threshold because
this is the span of the most extreme metallicity offset observed in
nearby mergers (e.g.\cite{kewley06, rupke08}).}
\label{fig:timehistograms}
\end{figure*}

The time of strong enhancement in the SFR also coincides with the
time of high-dilution.  Typically, we find that the dilution lasts for
$2\times10^9$ years, with half of the sample substaining a $z/z_{iso}\le
0.8$ for less than $5\times10^8$ (Fig.\ref{fig:timehistograms}). In the
case of fly-bys, this dilution is still visible far after the pericentre
passage, when the separation between the two galaxies is several hundreds
of kpc. We emphasise that the gas redistribution and subsequent
dilution takes place on scales of several kiloparsecs and is not due to
the effects of stellar feedback which typically act over much smaller
physical scales. Moreover, significant star formation is initiated {\sl
only} after a substantial gas inflow has occurred.  So feedback from
SNe explosion has little effect on the inflowing gas and certainly is
unable to prevent substantial inflows of gas initially.

A possible concurrent physical mechanism for redistributing metals in
the interstellar medium of the galaxy could be through gas diffusion,
which is not included in our code\footnote{We thank the referee for
pointing out the importance of diffusion in this problem.}.  At scales
of hundreds of parsecs (which corresponds roughly to our numerical
resolution), the diffusion law can take the form of a heat equation,
with a diffusion coefficient $D=9.25\times10^{26}cm^2s^{-1}$
\citep[see][]{martinez08}. This means that the typical
timescale for diffusing metals from the outer regions of the disk
($R=$7kpc\footnote{This is the average distance from the galaxy centre of
the gas that strongly participates in the interaction-driven inflow.}) is
$T_{diff}=R^2/D \ge 10^{10} yr$. This timescale is significantly longer
than the average inflow timescale which is of-order 300-400 Myr. Thus,
we can safely assert that the process of dilution of the central gas
metallicity studied here is driven by interaction-driven inflows rather
than metal diffusion.

Once the gas inflow and the corresponding burst of star formation is over,
the re-processed gas released by SNe explosions starts to enrich the
central metallicity, so that the $z/z_{iso}$ ratio becomes greater than
one (Figs. \ref{fig:timeevol} and \ref{fig:timeevol_fly}, fourth panel).
We checked that during the interaction, the contrast in the gas density
between the nuclear region and the outer disk are sufficient to result
in gas depletion times that are significantly lower than the times
typically required to re-enrich the ISM in metals.  This substantiates
our interpretation that the dilution in the metallicity of gas in the
nucleus is due to strong merger-driven gas inflows.

\subsection{Relationship Between the star formation and Metallicity}

\begin{figure*}

\begin{minipage}{0.9\textwidth}
  \centering
  \includegraphics[width=8cm,angle=270]{14304fg10.ps}
\end{minipage}\hfill
\centering
\includegraphics[width=4.cm,angle=0]{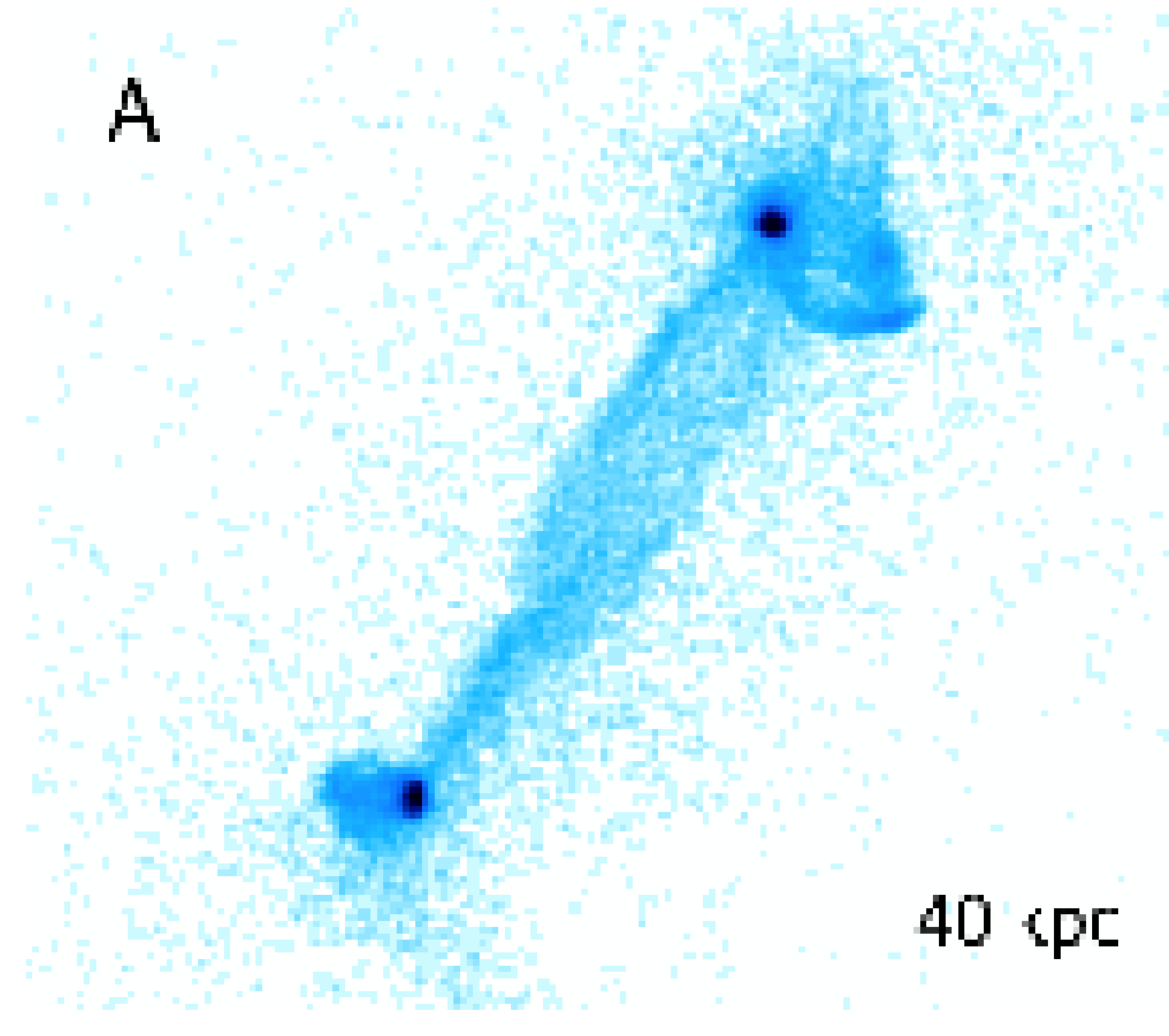}
\includegraphics[width=4.cm,angle=0]{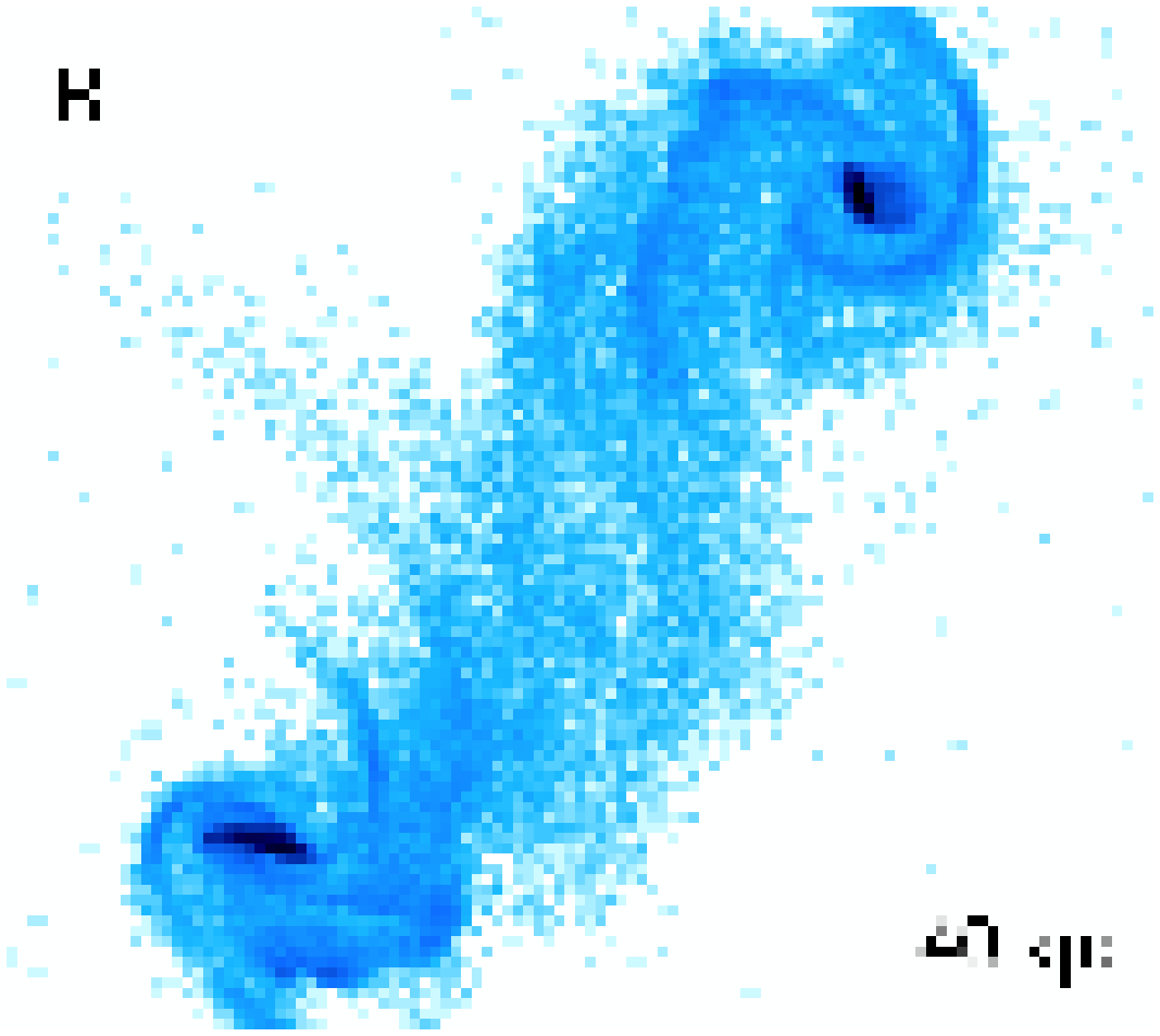}
\includegraphics[width=4.cm,angle=0]{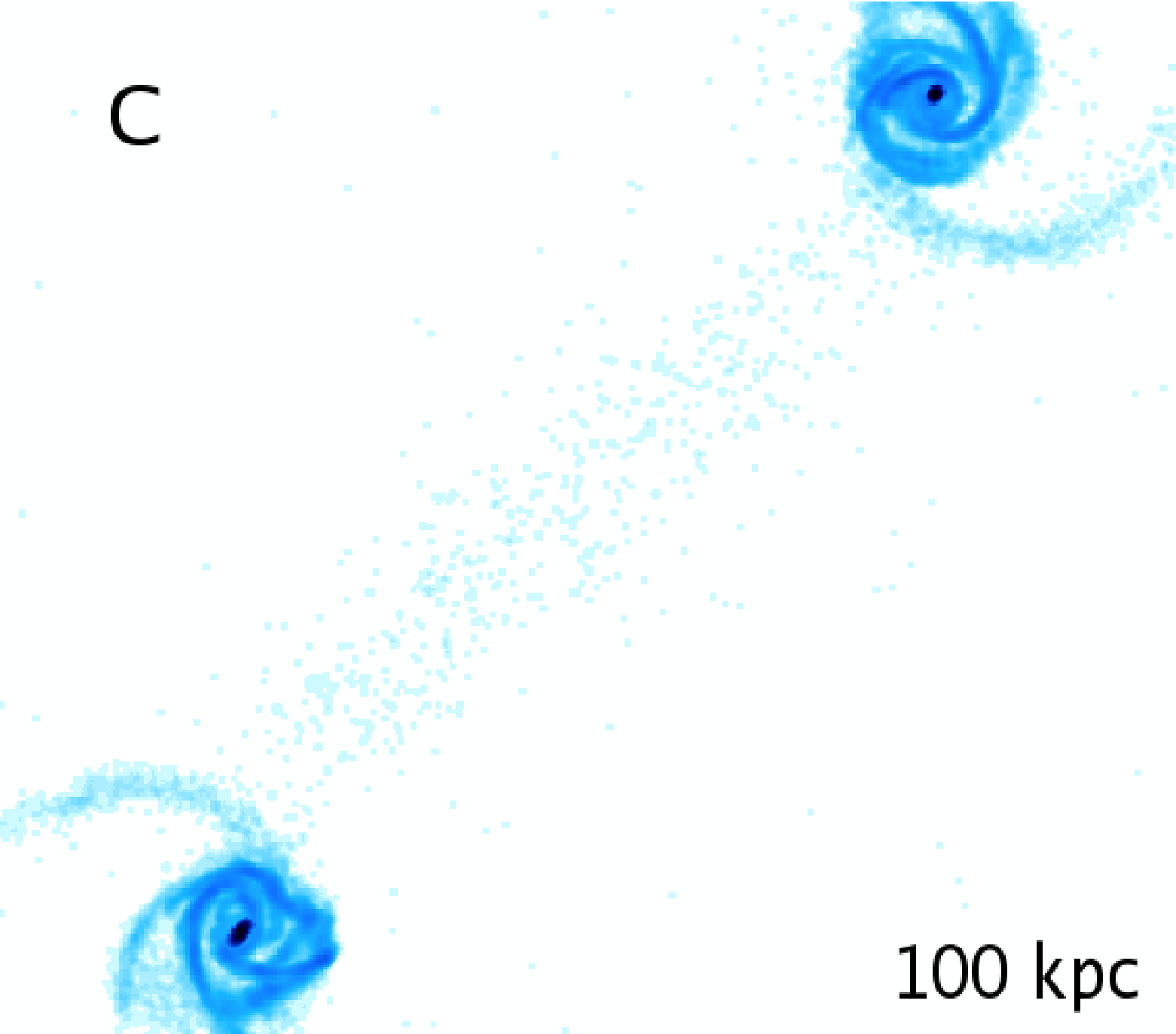}
\includegraphics[width=4.cm,angle=0]{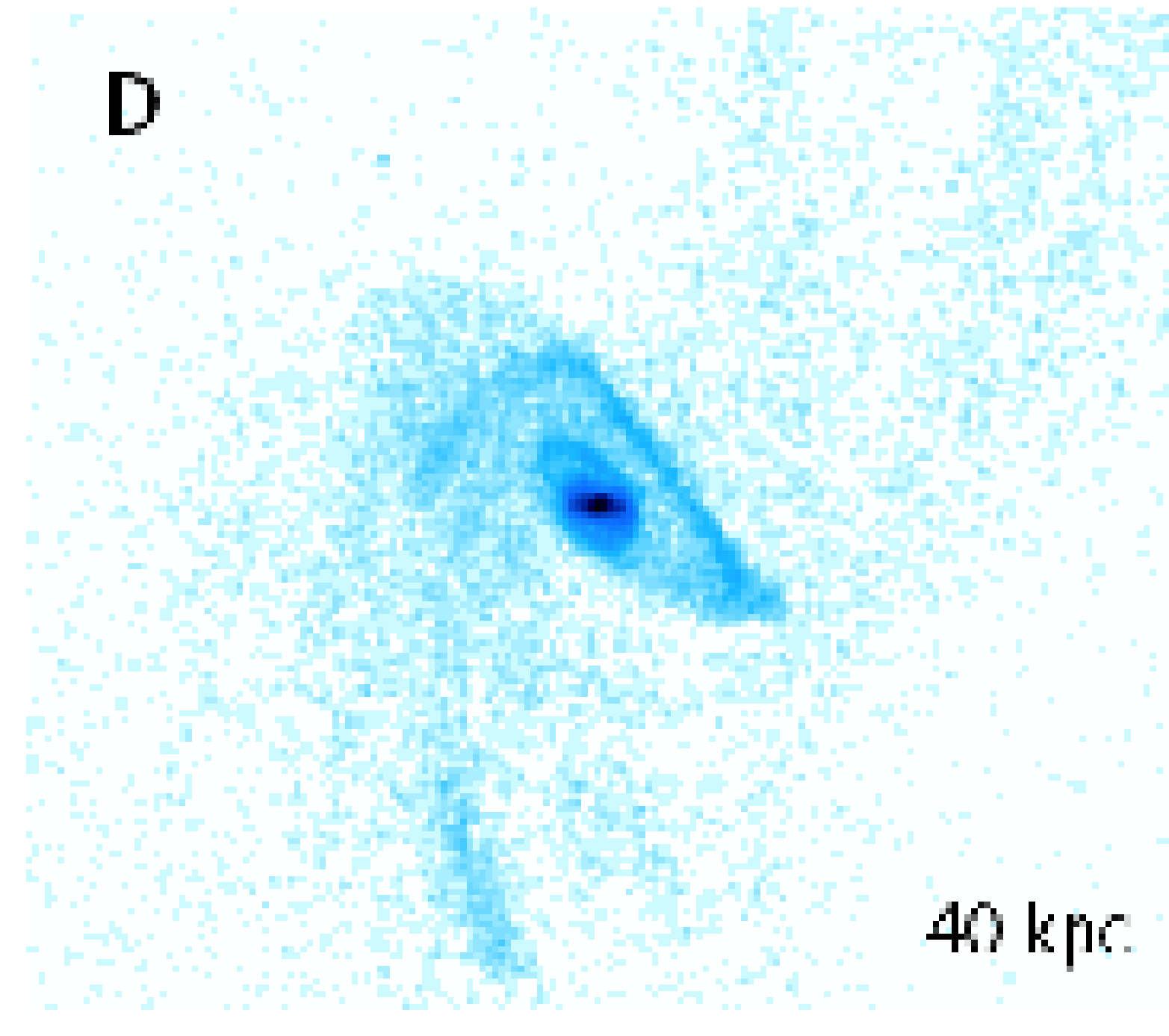}
\includegraphics[width=4.cm,angle=0]{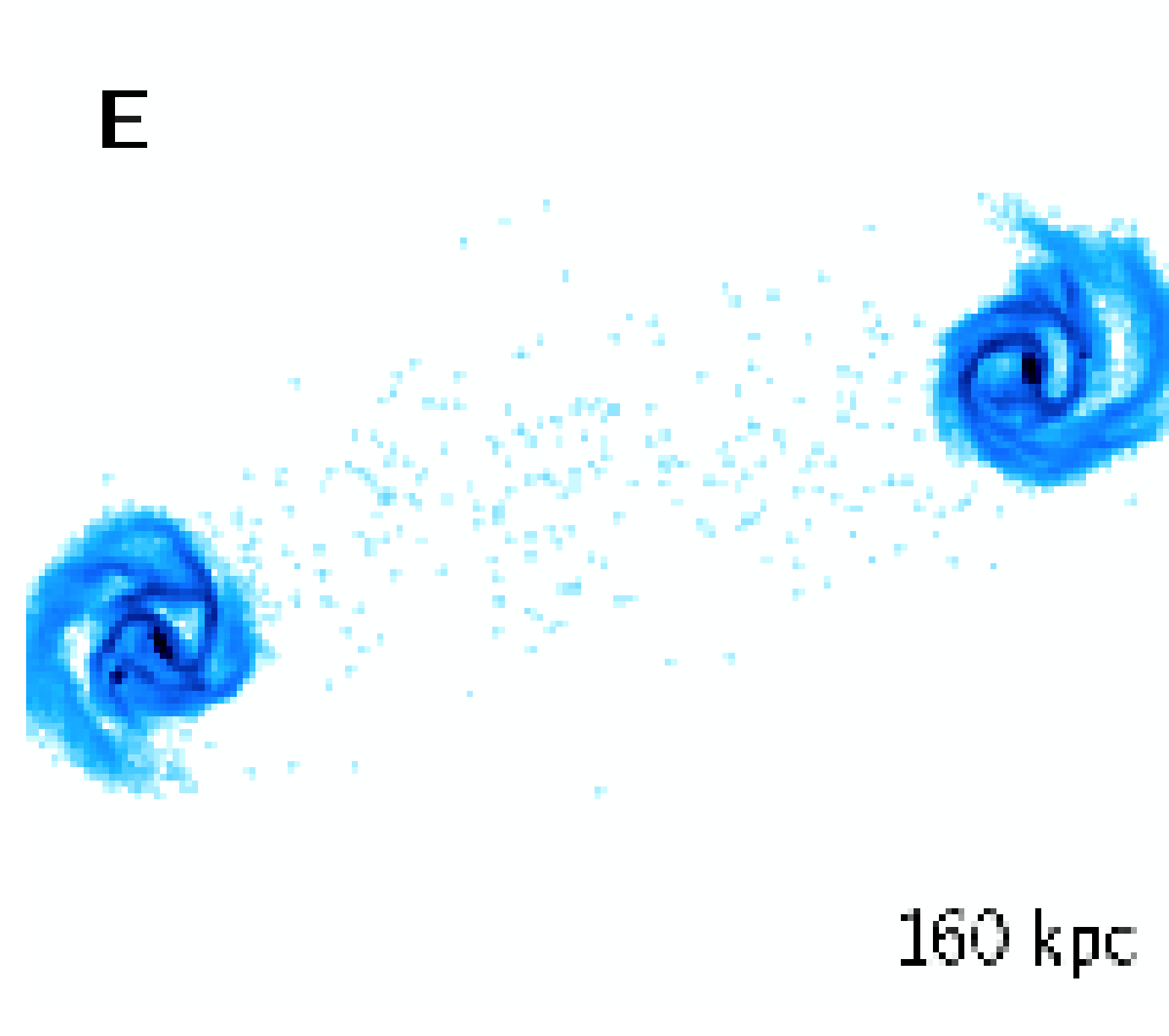}
\includegraphics[width=4.cm,angle=0]{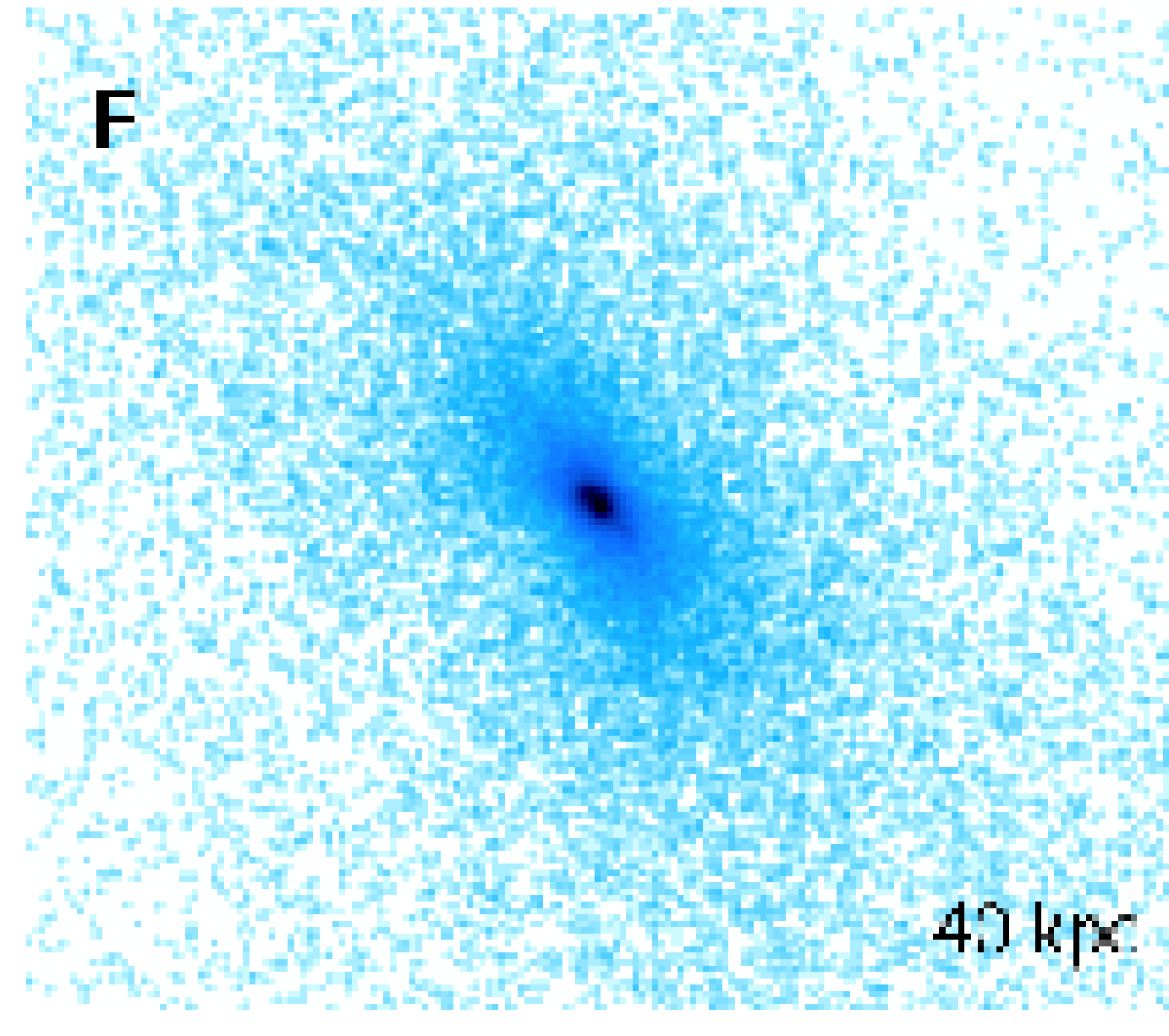}
\includegraphics[width=4.cm,angle=0]{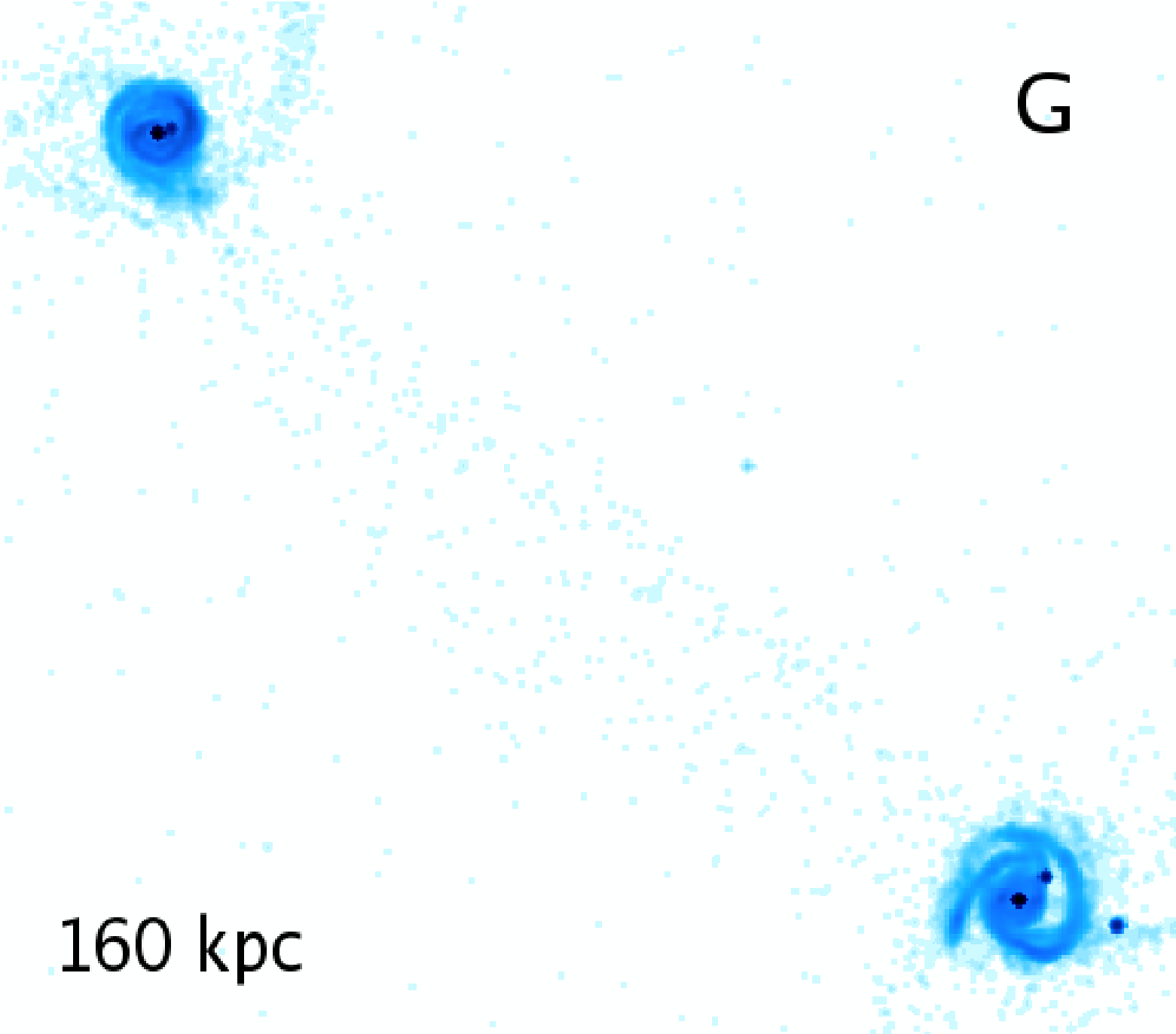}
\includegraphics[width=4.cm,angle=0]{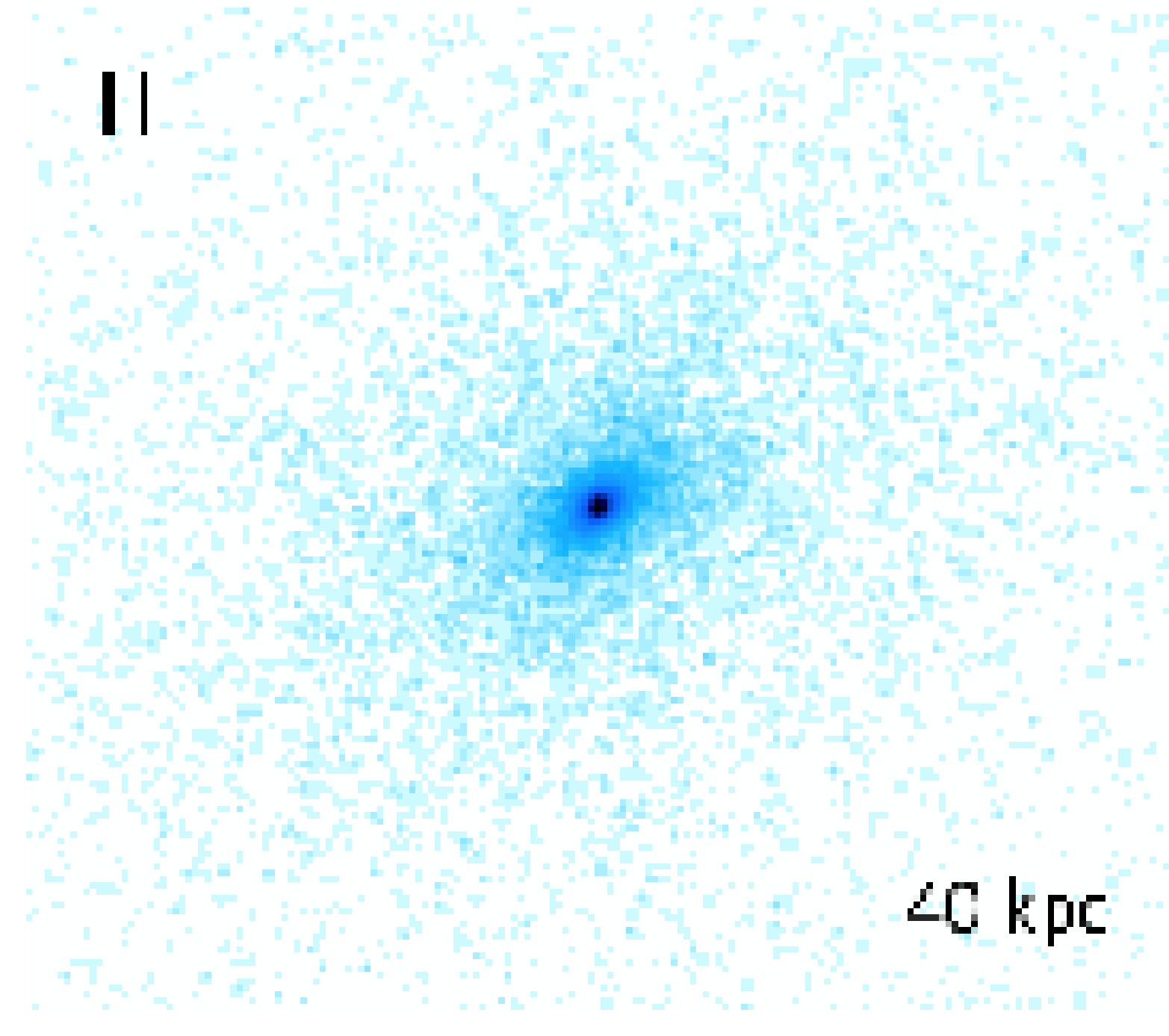}
\vspace{0.8cm}
\caption{The star formation rate, $\mathrm{M_{\sun}}$ $\mathrm{yr^{-1}}$, versus
the relative metallicity of the interaction and merger.  As in
Fig.~\ref{fig:timeevol}, the metallicity is shown relative to
an isolated disk with the same initial conditions as the galaxies used
in the merger simulations.  The blue asterisks indicate times when the
star formation is near its peak, while green triangles represent galaxies
before their peak star formation and red dots represent galaxies that
are past their peak of star formation and are in the final stages of
the merger. In the bottom panels, we show the morphologies of the merger
pairs in each region of the diagram to indicate the range of morphologies
seen in the simulations.}

\label{fig:sfrZrelation}
\end{figure*}

We have seen that the dilution of the gas-phase metallicity takes place
at approximately the same time as the strongest enhancement in the
star formation. In general, we find that the peak of SF and the maximum
dilution are roughly coeval, with a delay between the two of $10^8$ yr
at most.  This correlation between the starburst phase and the strong
dilution is clearly seen in the results of the simulations shown in
Fig.\ref{fig:sfrZrelation}.  The maximum dilution is reached in the
simulations at approximately the peak in the star formation rate. In
this phase, the higher SFR, which is a sign of the strongest gas inflow
are indeed associated with the strongest dilution as would be expected.
We note that a strong correlation between metallicity and central burst
strength has been observed by \citet{kewley06} in a sample of galaxy
pairs. Specifically, they pointed out that all five galaxies in the pairs
sample with strong central bursts have close companions and metallicities
lower than the comparable field galaxies.

Before the peak in the star formation rate, galaxies tend to populate
a region of low to moderate SFRs and dilutions (typically $0.7 \le
z/z_{iso}\le 1$; see Fig.\ref{fig:sfrZrelation}). 
During the burst phase and peak of
star formation rate, the ejecta of the SNe starts to enrich the gas, and
indeed, once the burst is over, galaxies have low star formation rates and
high-metallicities (Fig.\ref{fig:sfrZrelation}). 
 The first two phases of the relation
between star formation and dilution are associated with before, during,
and after the first pericentre passage and then close to final coalescence
of the merging galaxies, the last phase in the metallicity evolution is
always associated with well after their first pericentre passage and is
near or after coalescence.  This final state of the interaction is true
whether one considers mergers or simply flybys.

\subsection{Relationship between Separation and Metallicity}

\begin{figure*}
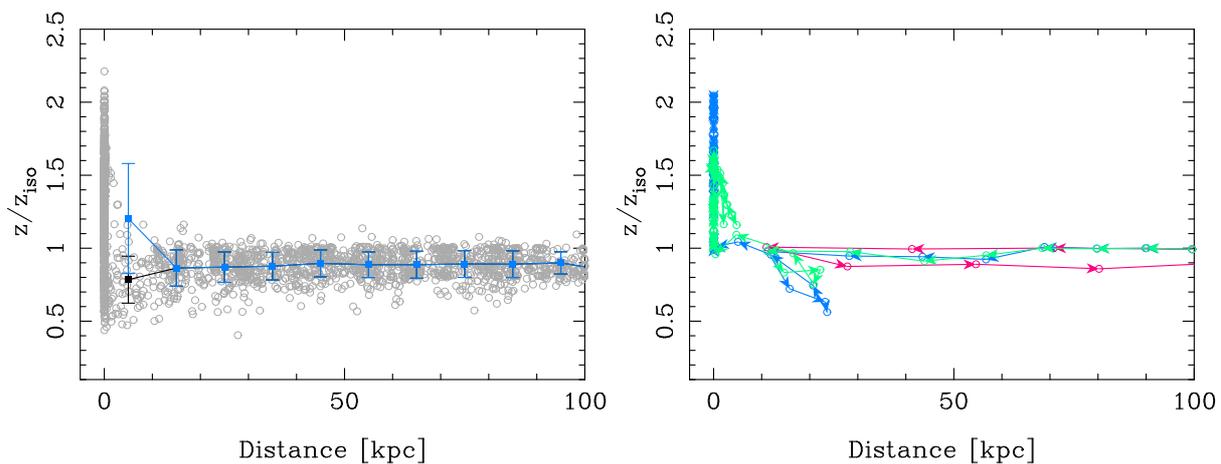

\centering
\includegraphics[width=6cm,angle=270]{14304fg19.ps}
\includegraphics[width=6cm,angle=270]{14304fg20.ps}
\vspace{0.8cm}
\caption{{\it (Left panel:)} The physical 3-dimensional separation
of the galaxy pairs and mergers versus their relative circumnuclear
metallicity. The black line represents the average relation, for physical
distances greater than 4 kpc and binned every 10 kpc. The blue line
represents the average relation when in addition including mergers with
physical separations below 4 kpc in the smallest bin (0-10 kpc). {\it
(Right panel:)} Evolution of two mergers (blue and green curves) and a
fly-by (red curve) in the metallicity-distance plane.  The simulations
suggest that even at wide separations there are gas inflows which
dilute the metallicities of the nuclear regions and that this dilution
persists until the pairs are strongly interacting, overlapping, and near
or after coalescence.  During the final merging process, a wide range
of metallicities can be observed but on average will show a much higher
metallicity than if they had not interacted or merged.}
\label{fig:sepZrelation} 
\end{figure*}

While the correlation between SFR (at the peak of intensity) and
metallicity dilution is quite remarkable, we expect a larger scatter when
plotting the dilution as a function of the physical distance of galaxies
in the pairs, especially for those at the smallest separations. This is
because the interacting pair can be in a number of different evolutionary
states at any given distance, except for perhaps the smallest separations.
For example, a galaxy at moderate separations of several 10s of kpc
could be in an initial encounter and before first pericentre passage
when the circumnuclear metallicity is largely unchanged.  Or it could
be just after the first pericentre passage when its circumnuclear gas could
be highly diluted (Fig.\ref{fig:sepZrelation}). This causes a range of
possible metallicity dilutions, from $z/z_{iso}=1$ to $z/z_{iso}\sim0.5$
(Fig.\ref{fig:sepZrelation}, right panel), with an average value of
$z/z_{iso}=0.75$. At the smallest separations, say less than 10-20 kpc,
but still distinguishable systems, galaxies are likely to be either
near the pericentre passage or near final coalescence.  This leads to
both a strong dilution and, on average, lower circumnuclear metallicity
(see Fig.\ref{fig:sepZrelation}).  At very small separations, less than
about 4 kpc, galaxies can be near first pericentre passage or near
coalescence which in either case have relatively large dilution and
relatively low-metallicity, or can have coalesced and be strong enriched
by supernovae ejecta and have relatively high-metallicity.  We show this
in Fig. \ref{fig:sepZrelation} where in the smallest separation bin,
if we include galaxy pairs with projected separations of less than 4
kpc we find a net average increase in the metallicity (i.e., $z>$1,
with an average value of $z/z_{iso} \sim 1.2$) while if we exclude
such systems, we find a net decrease in the circumnuclear metallicity
($z/z_{iso} \sim 0.7$).  This indicates the predominance of the metal
enriched mergers at small projected separations.

\citet{kewley06} have shown that galaxy pairs with small projected
separations ($4 \;\mathrm{kpc \; h^{-1}} \le s\le 20 \;\mathrm {kpc \; h^{-1}}$) and strong bursts of star
formation have metallicities lower than the comparable field galaxies.
Note that in this analysis we are considering the physical (3D) distance
between the two galaxies, and not their projected separation. Considering
the effects of projection would likely contribute to increasing the
scatter in the relation between metallicity and separation.

\subsection{O and Fe enrichment}

\begin{figure*}
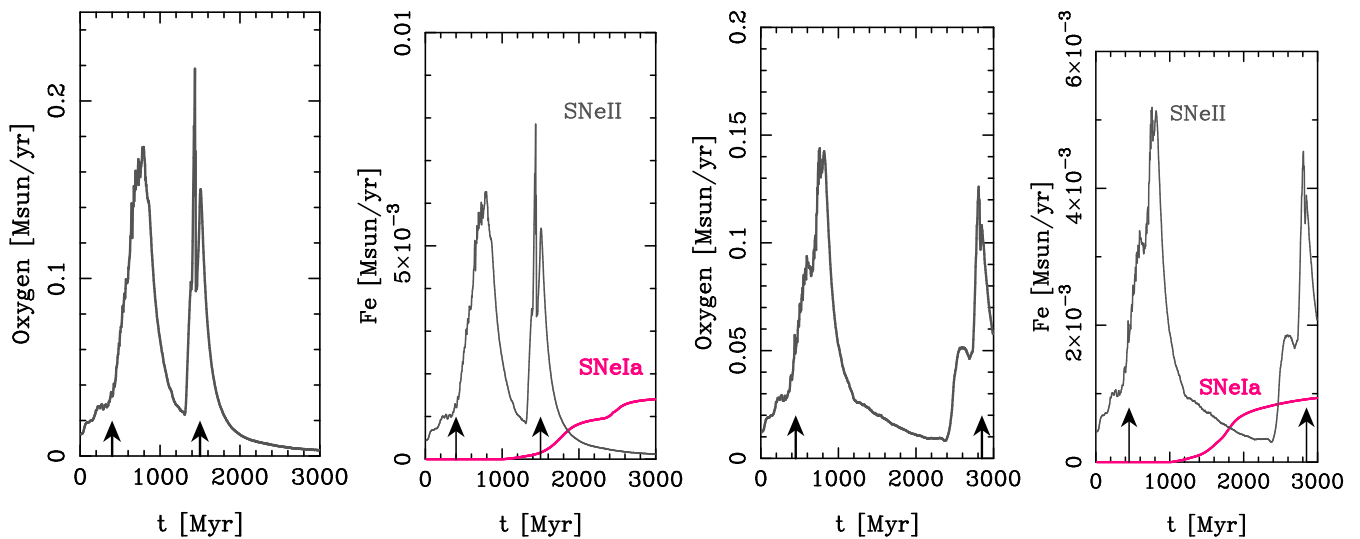

\centering
\includegraphics[width=7cm,angle=270]{14304fg21.ps}
\includegraphics[width=7cm,angle=270]{14304fg22.ps}
\includegraphics[width=7cm,angle=270]{14304fg23.ps}
\includegraphics[width=7cm,angle=270]{14304fg24.ps}
\vspace{0.8cm}
\caption{Rate of oxygen and iron released in the ISM by SNe II and
SNeIa explosions (in units of M$_{\sun}$ yr$^{-1}$) as a function
of time two different simulations whose star formation histories are
shown in Fig.\ref{fig:timeevol}. The first and final two panels show
the ejection rate of oxygen and iron for the star formation history
in Fig.\ref{fig:timeevol} represented by the orange and red curves
respectively.  For the iron release, we separate the contribution
from SNeIa (red curve) and SNe II (grey curve) for each of the two
simulations. The black arrows in each panel represent the time of the
first pericentre passage and final coalescence. Since the enrichment
is instantaneous (i.e., no delay for the evolutionary times of the type
II supernovae, the evolution of the star formation and ejecta rate are
synchronous.} \label{fig:oxygfe} \end{figure*}

\begin{figure*}
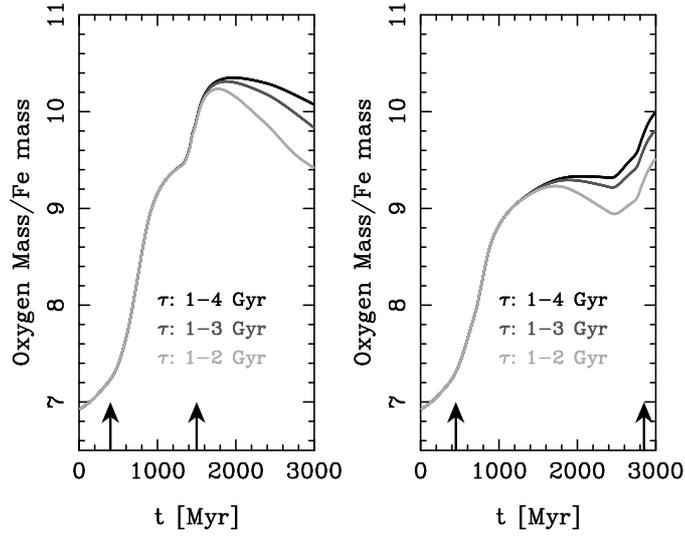

\centering
\includegraphics[width=7cm,angle=270]{14304fg25.ps}
\includegraphics[width=7cm,angle=270]{14304fg26.ps}
\vspace{0.8cm}
\caption{The evolution of the oxygen over iron mass ratio for two of
our galaxy mergers, whose rate of mass release in the ISM is shown in
Fig.\ref{fig:oxygfe}. For each panel, three different curves are shown
corresponding to three different delay times for onset of SNeIa, $\tau$:
$\tau=$1-4 Gyr (black curve), $\tau=$ 1-3 Gyr (dark grey) and $\tau=$ 1-2
Gyr (light grey), as indicated in the legends.  The black arrows in each
panel represent the time of first pericentre passage and coalescence.}
\label{fig:integrated}
\end{figure*}

Our simulations do not track the evolution of individual elements directly
in the code.  However, to track their evolution in an approximate way,
we can use some simple arguments to give an estimate of the amount of
iron and $\alpha$ elements, like oxygen, released in the ISM by type
I and II SNe.  At a given time, $t$, during the evolution of a galaxy,
the rate of oxygen mass released in the ISM by SNe II can be estimated by:

\begin{equation}
O_{SNeII}(t)[M_{\sun}/yr]=m_{O,SNeII}R_{SNeII}M_*(t)
\end{equation}

where $m_{O,SNeII}$ is the total mass of oxygen ejected averaged over
the initial stellar mass by a type II SNe, $R_{SNeII}$, is the number
of SNe II per unit mass of star formation, and $M_*(t)$ is the mass of
stars formed at time $t$.  Similarly, the rate of iron mass ejected on
average in the ISM by SNe II explosions is given by:

\begin{equation}
Fe_{SNeII}(t)[M_{\sun}/yr]=m_{Fe,SNeII}R_{SNeII}M_*(t).
\end{equation}

The progenitors of SNe II have very short lifetimes, less than a few
10 Myr, so that we can assume that the SNe II rate follows closely the
star formation rate of the galaxy, and so that the release of oxygen and
iron by SNe II in the ISM is approximately instantaneous.\footnote {Any
possible time delay would result into a time-shift between the SFR peak
and the oxygen and SNeII-Fe peak.}

The stars formed at time $t$ contributes also in enriching the ISM with
Fe released through type Ia SNe, which are the product of a thermonuclear
detonation of a white dwarf accreting matter from a companion star. While
the progenitors of SNe II have very short lifetimes, the time delay
between the birth of the progenitors of SNe Ia and their explosion can be
quite long, of the order of 1 Gyr or more depending on the progenitors,
the binary parameters, etc.\citep{strolg10}. We will assume delay time
between the birth of their progenitors and the explosion is uniformly
distributed between 1 and 4 Gyr after the birth of the parent population,
thus obtaining that, at any time $t$, the amount of iron released by
SNeIa is:

\begin{equation}
Fe_{SNeIa}(t)[M_{\sun}/yr]=m_{Fe,SNeIa}R_{SNeIa}M_*(t-\tau).
\end{equation}

where $\tau$ is the delay time between the birth of the progenitors
and the detonation of the type Ia supernova.  For the supernovae rates
$R_{SNeIa}$ and $R_{SNeII}$ we have used the values in  \citet{strolg10},
while the synthesised masses $m_{Fe,SNeIa}, m_{Fe,SNeII}$ and
$m_{O,SNeIa}$ are derived from Table 3 in \citet{iwamoto99}.  The amount
of oxygen released by a SNe Ia is about 0.1 times lower than the iron 
and we will neglect its contribution in the following estimations. In
Fig.\ref{fig:oxygfe},  we show the rate at which O and Fe are ejected into
the ISM during some of the simulated mergers. We caution the reader that
we are showing the global values, i.e. evaluated over the whole galaxy.
The intense episodes of star formation are accompanied by intense
peaks of O and Fe release, which coincide with the first phases of the
encounter, soon after the first pericentre passage, and with the merging
phase. During the burst of star formation, the ISM is enriched mostly
in oxygen (the $O_{SNeII}$ ejection rate is about a factor 30 greater
than that of $Fe_{SNeII}$). If a second burst of star formation occurs
in the simulation, a concomitant second peak of O and Fe release will
also take place in the final phases of the interaction causing further
relative enrichment of oxygen.

We note that if the coalescence of the two galaxies happens on a short
time scale (i.e., less than 1 Gyr after the first pericentre passage),
there is not enough time to enrich the ISM by SNe Ia. This means that the
stellar population forming during the final burst forms from gas still
relatively enhanced in $\alpha-$elements (see Fig.\ref{fig:oxygfe},
second panel). If, in turn, the merger happens on longer time scales
(i.e., longer than 1 Gyr), the iron released by SNe Ia explosions has time
to lower the [$\alpha/Fe$] ratio of the ISM, and the stellar population
formed in the merging phase will be less enhanced in $\alpha$ elements
(Fig.\ref{fig:oxygfe}, fourth panel).

The O and Fe ejected into the ISM obviously affects the ratio between
these two elements (Fig.\ref{fig:integrated}). Assuming that the
initial galaxies have percentages of O and Fe similar to the Sun, as
star formation proceeds during the interaction, the mass ratio of these
two elements changes: in particular the oxygen abundance increases,
until the iron release from SNeIa explosions inverts the trend. Note
that the influence of the iron release from SNe Ia on the populations of
new stars formed during the merger depends on the delay time between the
birth of the parent population and the onset of SNe Ia -- shorter delay
times cause a more rapid decline in the O/Fe ratio. We caution
the reader that in all our discussion of the relative abundance of
$\alpha-$elements, we are neglecting any possible contribution from SNe Ia
from the pre-existing old stellar population. This is of course an
important simplification of these models which we will investigate in
future studies.

\section{Discussion}

Overall, our results are not surprising.  It has been known for the last few
decades that mergers of gas-rich massive galaxies lead to transformations
in the morphological type \citep{barnes91, barnes92}  and drive strong
inflows of gas into the circumnuclear region \citep{casoli91,braine93,
barnes02, barnes96, dimatteo07}.  These inflows are a result of strong
tidal torques, dissipation, and strong asymmetries in the stellar
distribution  \citep[e.g., bars; see][]{barnes02, barnes96} leading to
efficient star formation \citep{dimatteo07}.  While the infall of gas
in mergers has been discussed within the context of simulations and
observations \citep{casoli91,braine93, barnes02, barnes96, dimatteo07},
currently, there has been only limited quantitative attempts to address
the relationship between infall, star formation, and the metallicity
\citep[e.g.,][]{perez06, rupke10}.  The trends we find in the simulations are
fascinating and explain a number of important phenomenon that have been
observed in interacting and merging galaxies.  Metallicity changes
during the merger event is a powerful way of testing the timing and
strength of the gas inflows and ultimately in understanding the fate of
post-merger/interacting galaxies.

\begin{figure*}
\centering
\includegraphics[width=8cm,angle=0]{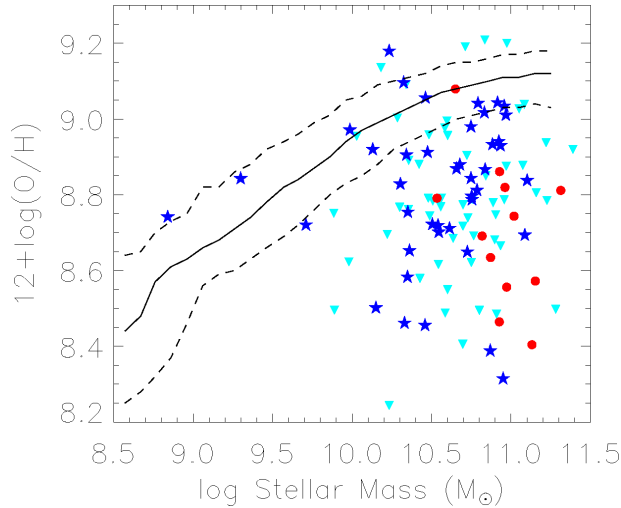}
\vspace{0.8cm}
\caption{Monte-carlo simulation of the metallicity distribution predicted
by the models compared to the data from \citet{rupke08}.  The up-side down
triangles are the simulated results (see text for details), the blue stars
and red circle are galaxy values taken directly from \citet{rupke08} as
described there.  Unfortunately, \citet{rupke08} did not actually measure
the mass of any of the individual galaxies but used a single mass
allowing for some scatter (see that paper for details).  We have approximately
reproduced their mass distribution for this comparison.}
\label{fig:MCmassmetallicity}
\end{figure*}

Overall, we find broad agreement with the results of our simulations
for the evolution of the metallicity and what is observed.  We find for
example that the net decrease in the metallicity is a few tenths of a dex,
which is comparable to what is observed in merging and interacting galaxies
\citep{kewley06,rupke08, dansac08, alonso10}.  In addition, as is likely
observed, our simulations show that the peak of the star formation
is close in time to the minimum in the circumnuclear metallicity
\citep{rupke08,rodrigues08}. The trends could be explained entirely by
the strength of the gas inflows of relatively metal-poor gas from the
outer regions of the disks which both fuel star formation and dilute
the metal content of the gas in and around the nucleus.  The strength
of the inflows likely depends both on the initial gas content of the
disks, how it is distributed, the characteristics of the orbits through
which the galaxies merge, and the mass ratios of the galaxies.  Here we
concentrated on the modelling a range of orbital parameters and a 20$\%$
gas fractions.  Subsequent papers will focus on varying the mass ratios,
gas fraction, and gas distributions.

The simulation results suggest that the strongest star formation and
dilution of the initial nuclear metallicity occurs around pericentre
passage of the merger.  This effect has been observed in that the
star formation at small separations exhibit a tail to high H$\alpha$
equivalent widths \citep[e.g.,][]{woods06, barton00}.  At this time,
we also expect the minimum in the metallicity to occur.  This expect
relationship has been emphasised in \citet{kewley06} based on a study
of pairs of galaxies.  The sample of \citet{kewley06} is particularly
appropriate for this type of study as it is a robust sample of pairs
of interacting galaxies and show generally a modest increase in their
star formation rates and modest likely dilutions.  For example,  their
most intense starbursts, which show significant offsets in metallicity,
only have offsets of about 0.2-0.3 dex compared to non-interacting
galaxies with similar B-band absolute magnitudes.  This is the amount
we find in our simulations.

We can explain the full range of circumnuclear metallicities observed
in more violent mergers in the local \citep{rupke08} or distant universe
\citep{rodrigues08}.  By violent mergers, we mean in this context galaxies
which are luminous in the infrared and have star formation rates several
to almost 100 times that of the Milky Way.  These are the most appropriate
for comparison because, as we have shown, as the merger advances, the
star formation rate will decrease and the metallicity increases reaching
values that are higher than the initial metallicities.

To show that our models can reproduce the data, we have made a
monte-carlo simulation of the metallicity distribution we might obtain
if we observed a set of galaxies near their maximum star formation rate
and allow for a range in their properties and the intrinsic scatter
in the mass-metallicity relationship.  For such a comparison sample,
we would naturally draw on the most violent star-forming galaxies in
the local universe which are major mergers.  This is the point in our
simulations, where we get the most significant influx of material into
the circumnuclear regions and near the peak of star formation.  At this
stage, we can dilute the circumnuclear gas with relatively metal poor
gas from the outer disk decreasing the overall metallicity by about a
factor of 2 and relatively brief time (of-order an internal dynamical time
of about 10$^8$ yrs).  Thus, in our simple monte-carlo approach to the
metallicity estimates, we allow for a decrease of 0.3$\pm$0.1 dex. We make
this allowance as we only have single metallicity gradient in our all our
models ($-$0.1 dex kpc$^{-1}$).  However, metallicity gradients in spiral
galaxies range from about 0.0 to $-$0.2 dex kpc$^{-1}$. To allow for this
variation, we have assigned a range in the metallicity allowed of 0.1 dex.
We do not know the nature of the progenitors either.  They could have a
range of initial circumnuclear metallicities and the mass metallicity
relationship has a significant scatter 0.15 dex \citep{tremonti04}.
Thus we allow for an additional source of scatter of 0.15 dex to reflect
the scatter in the mass-metallicity relation.  Finally, there is also
measurement uncertainties which are about 0.1 dex \citep{rupke08}.

We assume that each source of scatter is Gaussian and that each
can be linearly combined.  This assumption is appropriate since (at the metallicities
observed here) the uncertainty in the metallicity is not correlated
to the metallicity, our numerical simulations are completely
independent of the data, and the scatter in the mass-metallicity
relationship is intrinsic.  One such monte-carlo simulation is shown
in Fig.~\ref{fig:MCmassmetallicity} but all reproduce the data from
\citet{rupke08} reasonably well.  Unfortunately, \citet{rupke08} did not
actually estimate masses directly for their sample but used a random set
of masses generated from a typical mass and scatter from the literature
 where then assigned.  While we could have used their masses directly
which would have made the agreement look more spectacular, the reality is
that we have simply reproduced the scatter in the metallicity by making
a set of simple assumptions.  Thus this agreement between the data and
model should not be taken literally but demonstrates that numerical
simulations of the type run here can reproduce the spread in metallicity
without making any assumptions beyond suggesting that the highest levels
of star formation in a merger occurs near the peak in the strength of
the gas flows something that every model over the last decades has shown
and can be argued on robust physical grounds.

In addition to investigating the general evolution of the metallicity,
we also adopted a crude model for the evolution of Oxygen and Iron during
mergers and interactions.  We were motivated to make these simplistic
estimates by the fact that following the evolution of $\alpha$ elements
like Oxygen can be used to  ``date'' mergers.   In analogy with
studies of the stellar populations of early type galaxies, bulges,
or in components of the Milky Way such as the thick disk and bulge,
[O/Fe] or [$\alpha$/Fe] ratios can be used as a chronometer telling
us over what timescale the bulk of the (luminosity-weighted) stellar
population formed \citep[e.g.,][]{thomas05, zoccali06}.  Our hope was
that even a crude model such as the one we have adopted could provide
useful information about how observationally we might go about dating
mergers and interactions.

How this might work is that if the galaxy is in the early phases of
merging, we would expect to see both overall low-metallicity and
depending on the distribution of [$\alpha$/Fe] in initial disks,
evidence of what radii the gas came from through its abundance ratios.
The bulge of our galaxies, at least in the stars, is enriched in
$\alpha$ elements relative to the sun \citep{lecurer07, zoccali08,
ryde10}. Moreover, the disk of our Milky Way appears to have an enhancement
out to relatively large radii but this ratio depends on the radius
\citep[see][and references therein]{acharova10}.  Of course these
measurements are for the stellar metal abundances, not for the gas
phase so the comparison is not completely fair. It does however at least
suggests a possible trend that may be useful in determining the radii at
which freshly deposited gas came from.  At any rate, we may therefore
expect in the early stages, a mild $\alpha$ enhancement in the gas.
As the merger proceeds, this enhancement will become more pronounced,
until the late stages where it will begin to lessen as the type Ia SNe
become more important.  An interesting question is how much of this
enhanced material makes its way into each generation of star formation.
While in principle this can happen quickly as the star formation lasts
for several 100 Myrs (several crossing times) sufficient for the metals
to mix with the ambient gas, it is not clear if this actually happens.
If it does, then we would predict that a majority of the stars formed
would be $\alpha$ enhanced but may have metallicities lower than stars
formed from gas before the burst occurred.  Only after about 1-2 Gyr
or so would both the metallicity increase above what an isolated galaxy
would have and would the $\alpha$ enhancement begin to lessen.  Of course
this exact time scale is uncertain for both the metallicity and  $\alpha$
enhance because we do not have a firm understanding of the delay time
for type Ia SNe.

Taken at face value, the results of our simulations would suggest that
we can make realistic bulges and early-type galaxies.  It is well known
that the bulge of the Milky Way is both metal-rich [typically about
[Fe/H]$\sim$-0.3 to 0.3 with a mean of about 0]\citep{zoccali08} and is
$\alpha$ enhanced \citep{lecurer07, zoccali08, ryde10}.  Both of these
results are consistent with our simulations modulo the uncertainty
in the delay times of type Ia SNe, and the mass and characteristics
of pre-existing bulges in the merging galaxies\citep[see][]{pipino09}. 
This general result appears to
hold generally for bulges, namely they are metal-rich and $\alpha$
enhanced \citep{thomas05} and normalizing for their mass differences,
very similar to early type galaxies \citep{thomas05}.
Interestingly, there is some evidence for a radial gradient in the
average metallicity of the stars in the galactic bulge \citep{zoccali08}.
Depending on the initial conditions, this may be consistent with our
simulations.  In the simulations it is likely there is a gradient in
the star formation with radius as the strongest gas flows and most
intense star formation is in the deepest part of the potential well.
However, a complete discussion of the metallicity, $\alpha$ enhancement,
and radial gradients in bulges and early-type galaxies is beyond the
scope of this paper, our simulation results are at least consistent with
these general properties of spheroids.  We will address these questions
in more detail in a subsequent manuscript.

Finally, these results have important implications for our understanding
of metallicity evolution of galaxies as observed at high-redshift.  Since
we have found that during the most intense phases of star formation, the
gas phase metallicity can be reduced (diluted) by infalling material from
the outer regions of the merging galaxies.  Mergers and interactions are
known to play an important role in initiating intense starbursts in the
distant universe \citep{law07, nesvadba08, law09, epinat09}.  In order to
estimate gas phase metallicities of distant galaxies, they must have high
surface brightnesses in their emission line gas \citep[see][and references
therein ]{lehnert09}.  It is now reasonably well established that the
mass-metallicity relation appears to have evolved over the last 10 Gyrs
\citep{savaglio05,erb06,mannucci09, rodrigues08, halliday08, queyrel09}.
However, combining the possible limitations in the observations due to
surface brightness dimming and the importance of mergers in initiating
intense (high surface brightness) starbursts, it may be the determinations
of the evolution in metallicity may be strongly influenced by dilution.
If this is so, we would predict that, if the importance of mergers
increases with redshift \citep[e.g.,][]{deravel09}, we should see both
a decrease in the average metallicity and an increase in the scatter
in the mass-metallicity relation with increasing redshift.  While we
have no modelled a full range of initial conditions, our results here
suggest that this increasing scatter would be about several tenths of
a dex which is less than a factor of 2 more than the scatter in the
relation for local galaxies \citep{gallazi05}.

\section{Conclusions}

In this work, we have studied the dilution and subsequent enrichment
of the interstellar medium during interactions and mergers of disk
galaxies of comparable mass.  Our models, which include star formation
and chemical enrichment, have made it possible to follow, for the first time,
the evolution of the circumnuclear gas phase metallicity from the initial
interaction until the coalescence phase, thus significantly improving
the existing simulations and leading to insight between star formation, gas
inflows and dilution, and metallicity evolution in mergers and interactions.

Our main results can be summarised as follows:

\begin{itemize}

\item The dilution of the gas initially in circumnuclear regions of
the merger pairs starts just after the first pericentre passage, as
the disk is destabilised and gas flows inwards. For mergers, usually a
second dilution peak is observed in the final phase of coalescence. A
significative dilution can take place also in fly-bys, just after their
close passage, mimicking to some extent what we observe in our merger
simulations.  Thus even flybys may be important for the star formation
and evolution of galaxies.

\item On average, the amplitude of the metallicity dilution we find
(0.2-0.3 dex) is in agreement with observations \citep{kewley06,rupke08}.

\item The strongest correlation we found is between the maximum in the SFR
and the strength of the circumnuclear dilution -- pairs experiencing the
strongest bursts also show the strongest circumnuclear dilution. This
can be explained in terms of interaction-driven gas inflows from the
outer disk regions into the galaxy centre.

\item The dilution phase (defined as $z/z_{iso} <$0.8) is very rapid,
half of the sample sustains this high dilution phase for less than
$5\times 10^8$ years.

\item The gas inflow causes concomitant
enhanced star formation; such enhanced star formation releases a
significant amount of enriched material in the ISM. This enrichment
results in merger remnants which have higher final circumnuclear
metallicities than the corresponding galaxies evolved in isolation. The
same can happen for fly-bys which experience enhanced star formation
rates.

\item Type Ia and II SNe formed during the interaction release O and Fe in
the ISM, thus changing the average mass ratio of these two elements. In
particular, we note that, if the merger takes place on short time scales
(lower than the typical delay time of SNe Ia, which unfortunately is not
well constrained), the stellar population formed during the last SF burst
forms from an $\alpha-$enriched ISM, while, for mergers with longer time
scales, SNeIa can have the time to release a sufficient amount of iron
in the gas. The resulting stellar population has thus lower values of
$[\alpha/Fe]$ ratios.

\item The circumnuclear metallicity, as well as the
[$\alpha/Fe$] ratios, can be used as an indicator of the timing of merger/interaction
state, thus helping in disentangle projection effects, i.e., galaxies
which appears close in projection, but that are not strongly interacting.
The remnants of these mergers appear qualitatively consistent with what is
known about the bulge of our galaxy, other nearby galaxies, and early
type galaxies generally.  Thus, in agreement with many authors, it appears
when considering chemical evolution that mergers can reproduce early type
systems.

\item If mergers play a significant role in the galaxies that have had their
metallicities determined at high-redshift, then it is possible that dilution
is signficantly affecting these estimates.

\end{itemize}

\section*{Acknowledgments}
PDM is supported by a grant from the French Agence Nationale de la
Recherche (ANR). These simulations are available as part of the GalMer
simulation data base (\emph{http://galmer.obspm.fr}). The authors wish
to thank the referee for their constructive and helpful comments and
A. Pipino for a critical reading of an early version of this manuscript
and helpful suggestions.

\begin{appendix}

\begin{figure}
\centering
\includegraphics[width=6cm,angle=270]{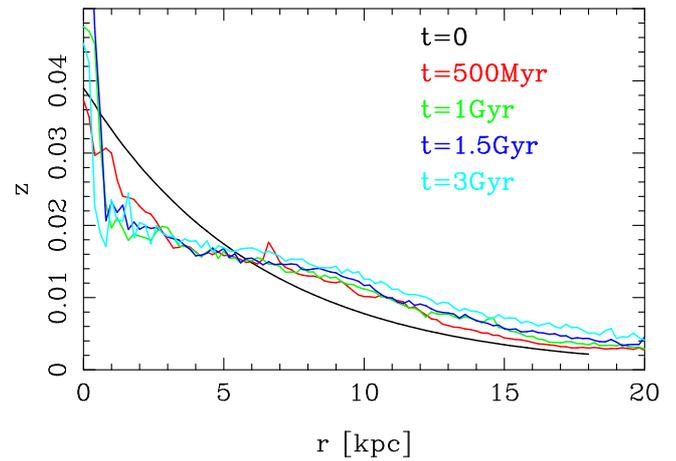}
\caption{Metallicity profiles of the gas in the gSb galaxy simulation
which were evolved in isolation (i.e., did not experience a merger or
interaction during the simulation). The profiles are given at different
times, as indicated in the labels. }
\label{fig:isoevol}
\end{figure}

\begin{figure}
\centering
\includegraphics[width=2cm,angle=0]{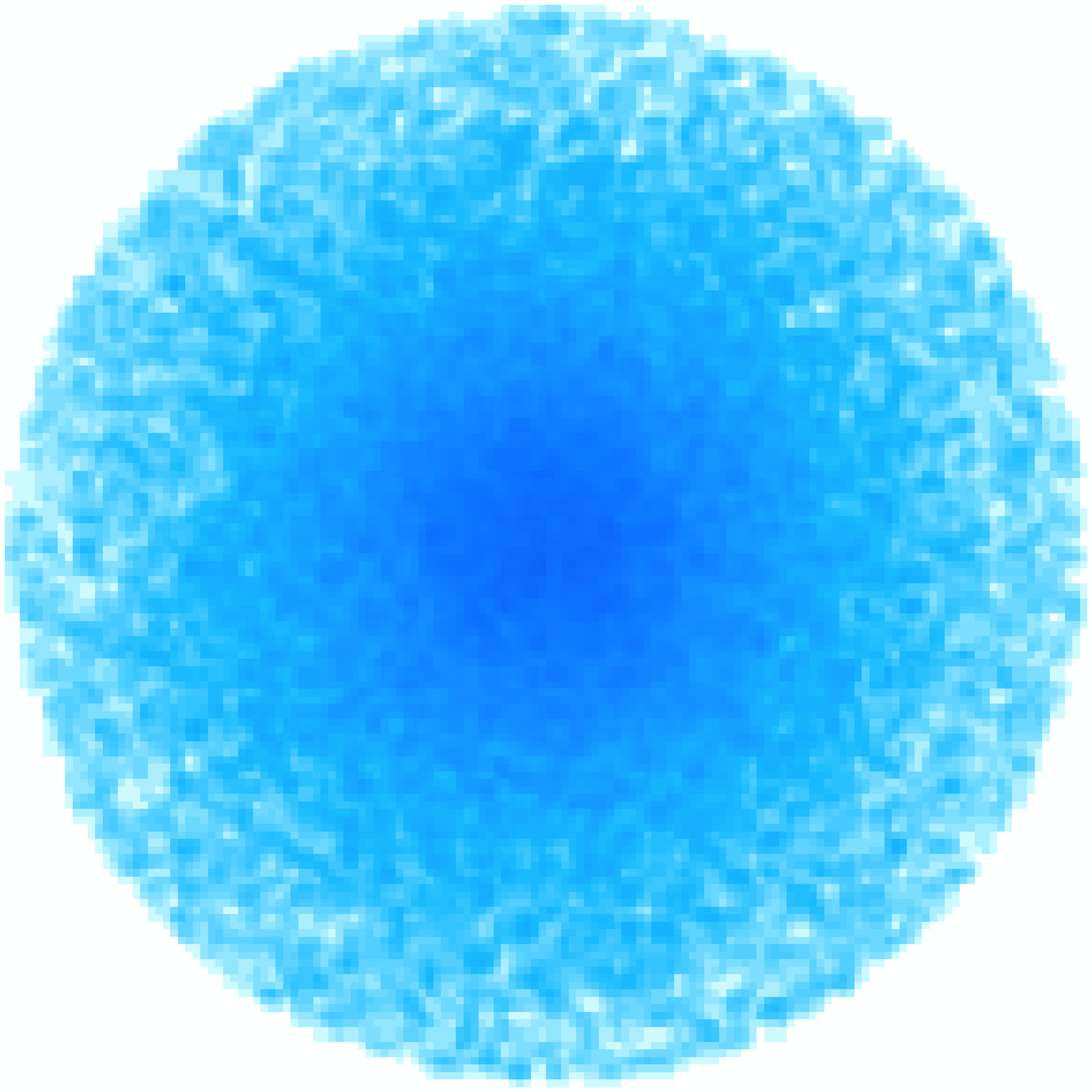}
\includegraphics[width=2cm,angle=0]{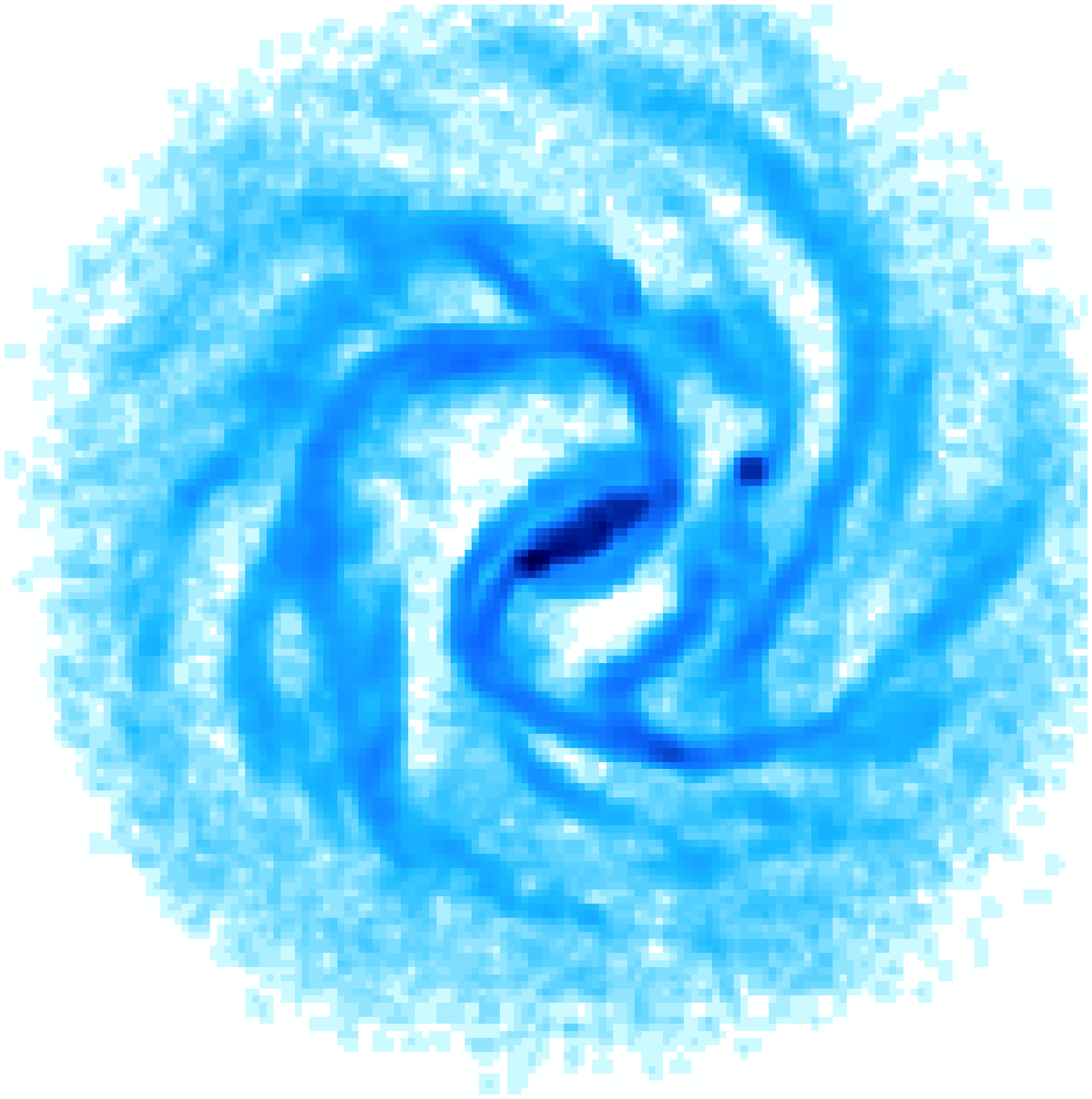}
\includegraphics[width=2cm,angle=0]{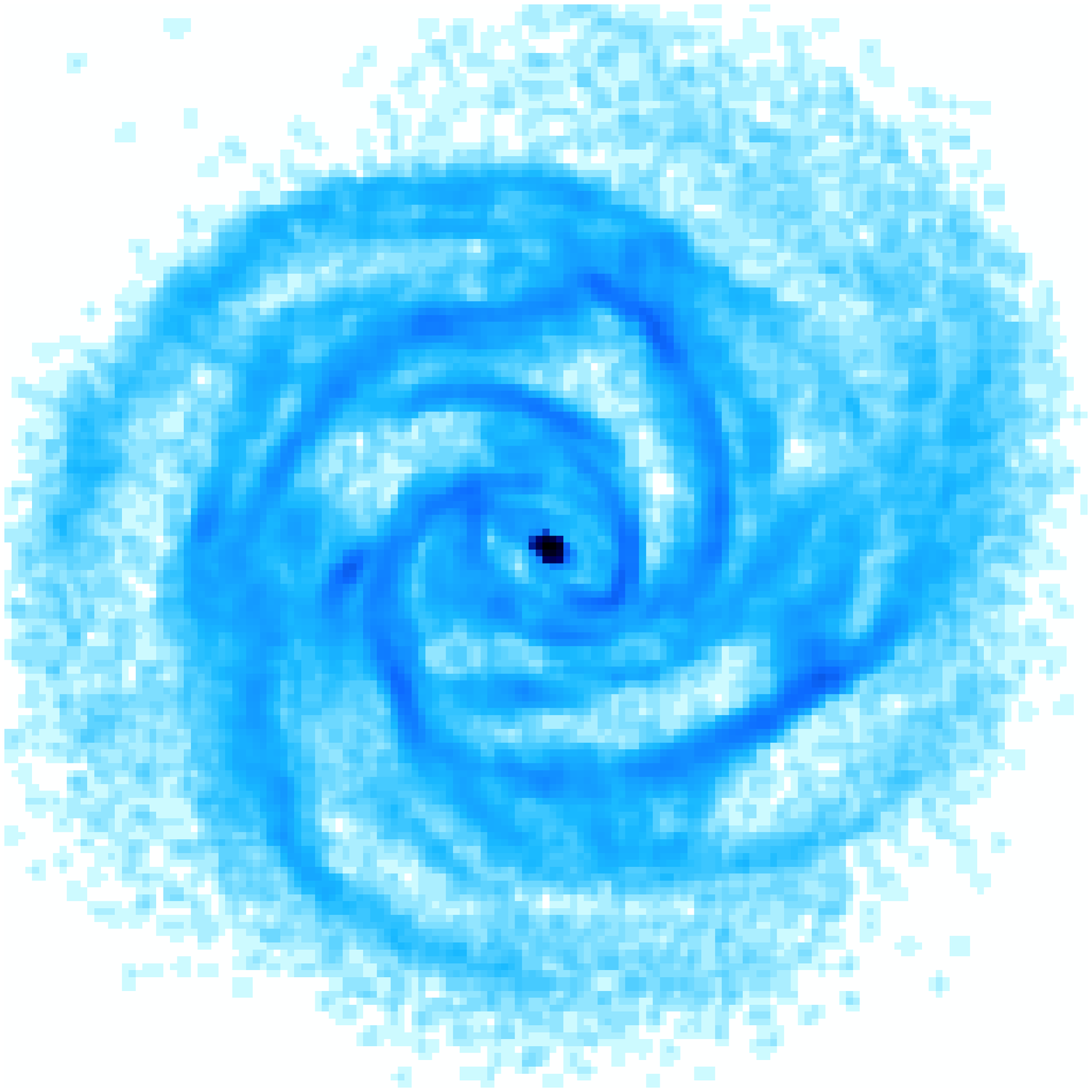}
\includegraphics[width=2cm,angle=0]{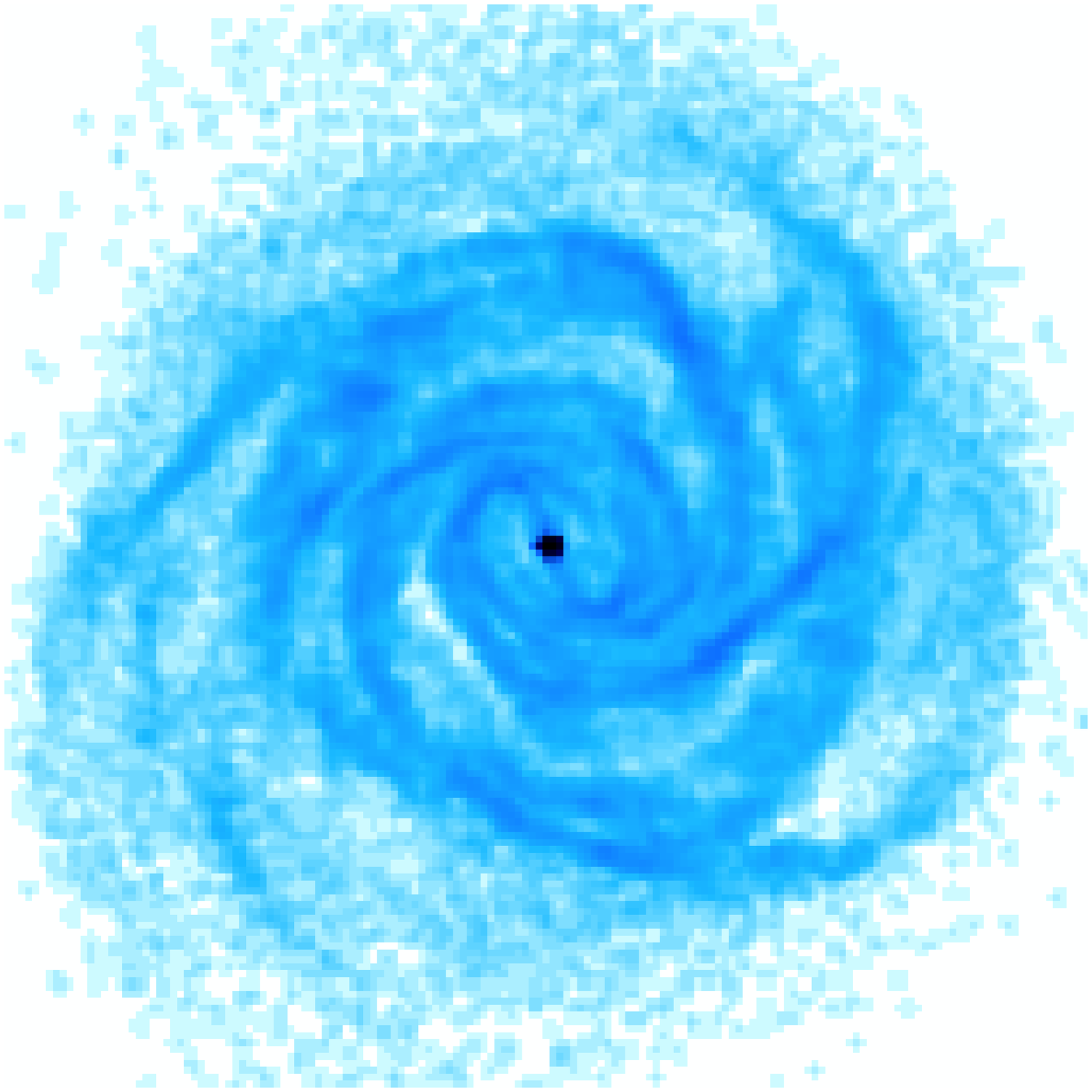}
\includegraphics[width=2cm,angle=0]{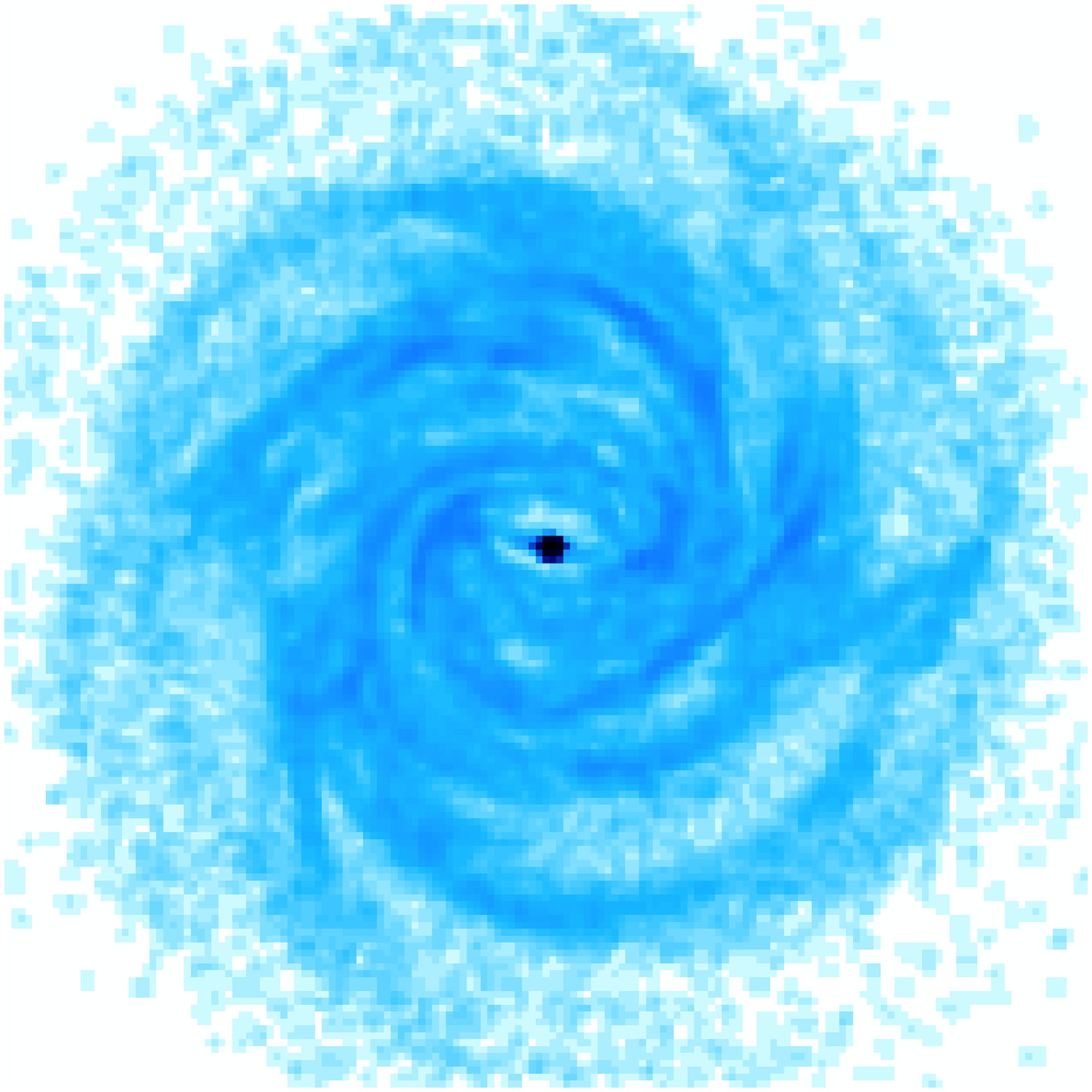}
\includegraphics[width=2cm,angle=0]{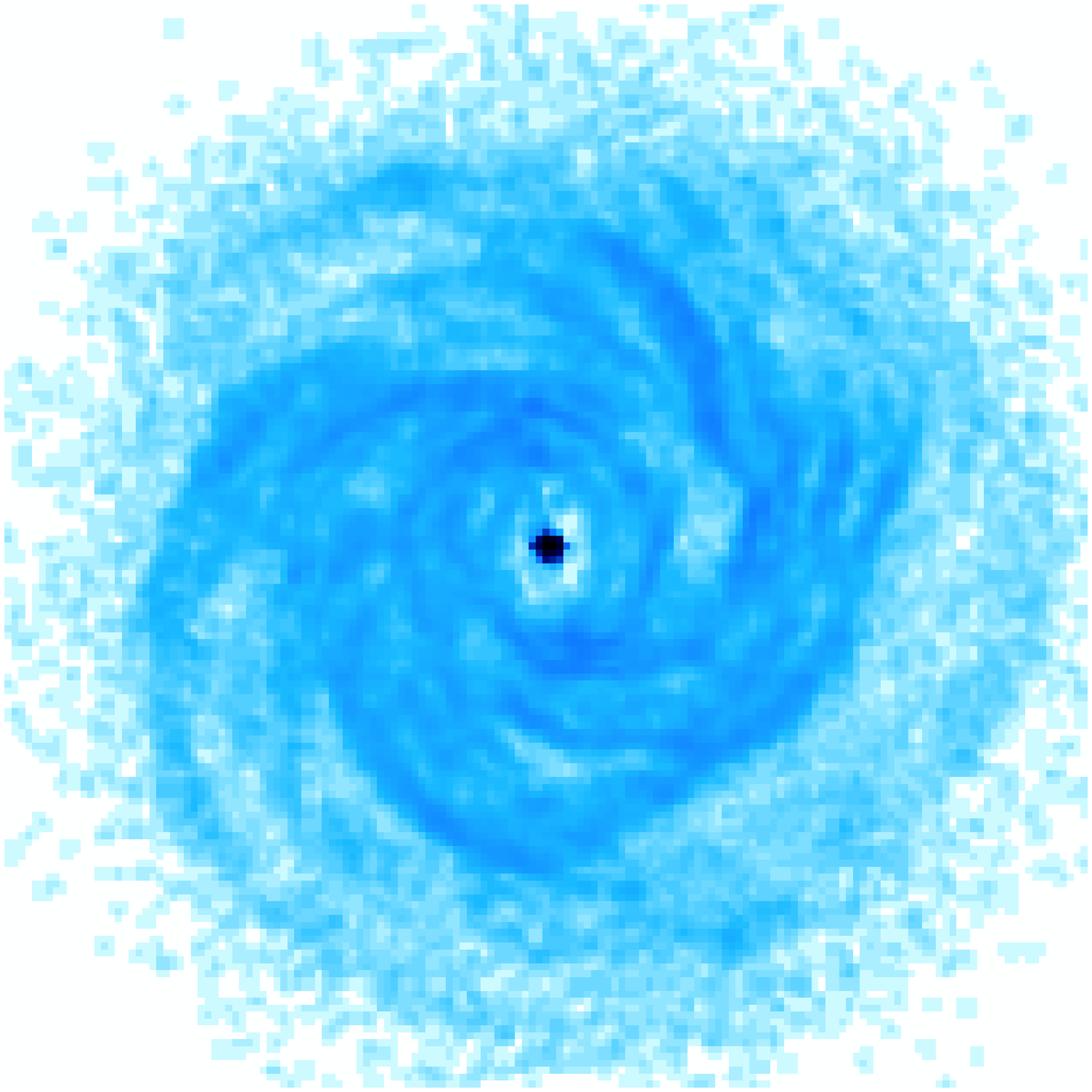}
\includegraphics[width=2cm,angle=0]{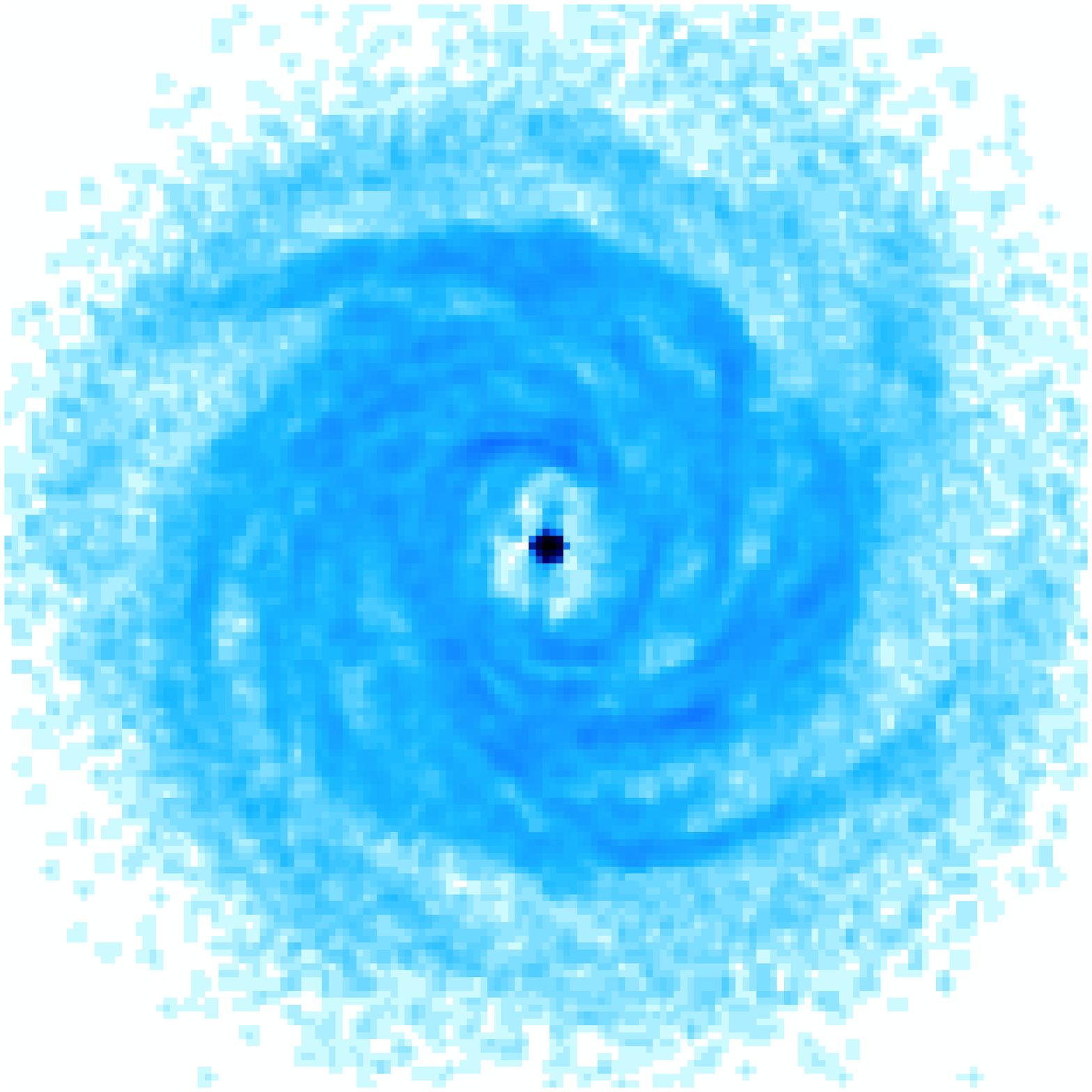}
\caption{From top-left to bottom-right: Maps of the gas distribution of the model gSb galaxy evolved in isolation. Maps are shown from t=0 to t=3 Gyr in steps of 500 Myrs. Each map is 40 kpc $\times$ 40 kpc in size.}
\label{fig:isogas}
\end{figure}

\begin{figure}
\centering
\includegraphics[width=2cm,angle=0]{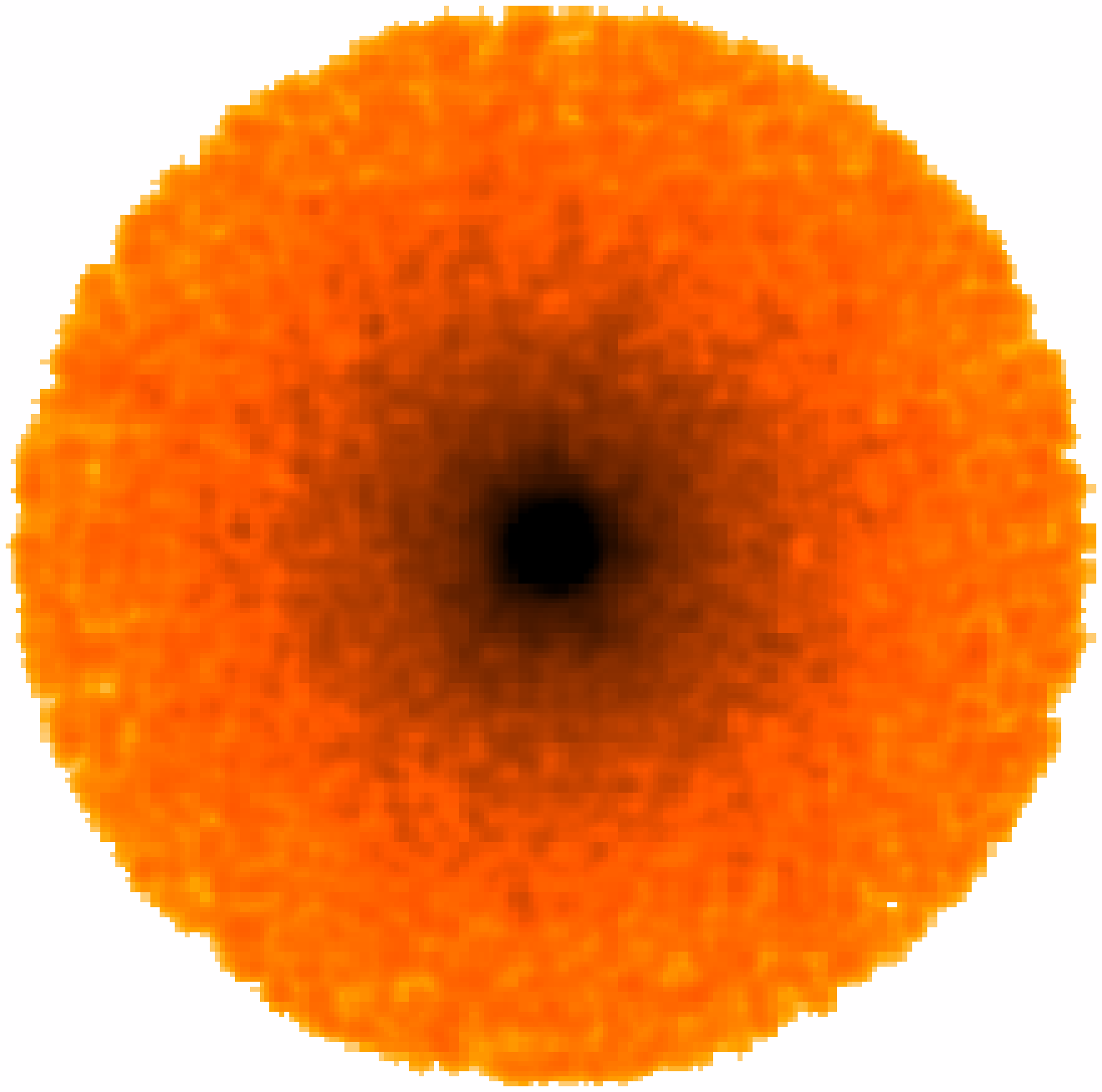}
\includegraphics[width=2cm,angle=0]{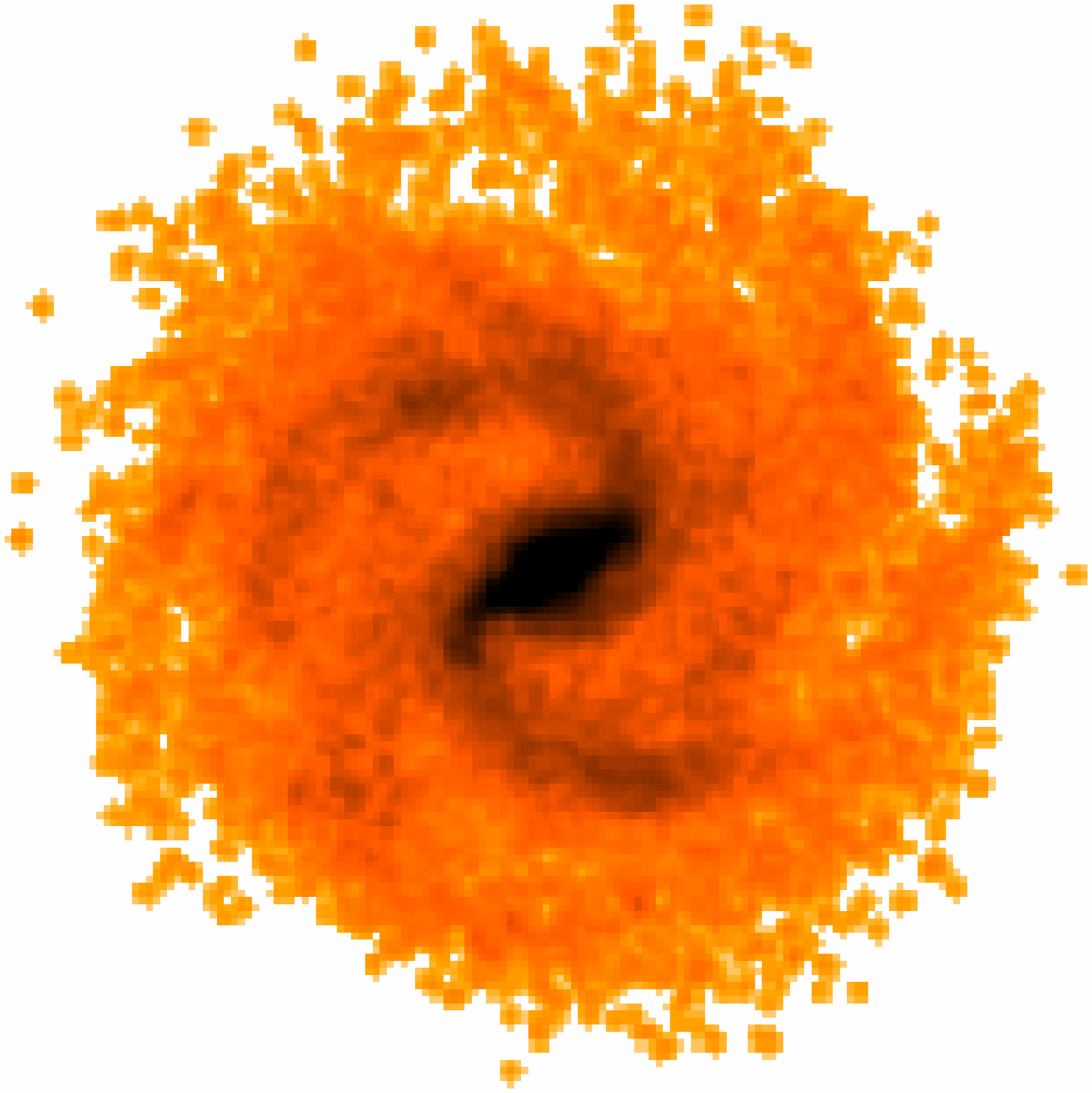}
\includegraphics[width=2cm,angle=0]{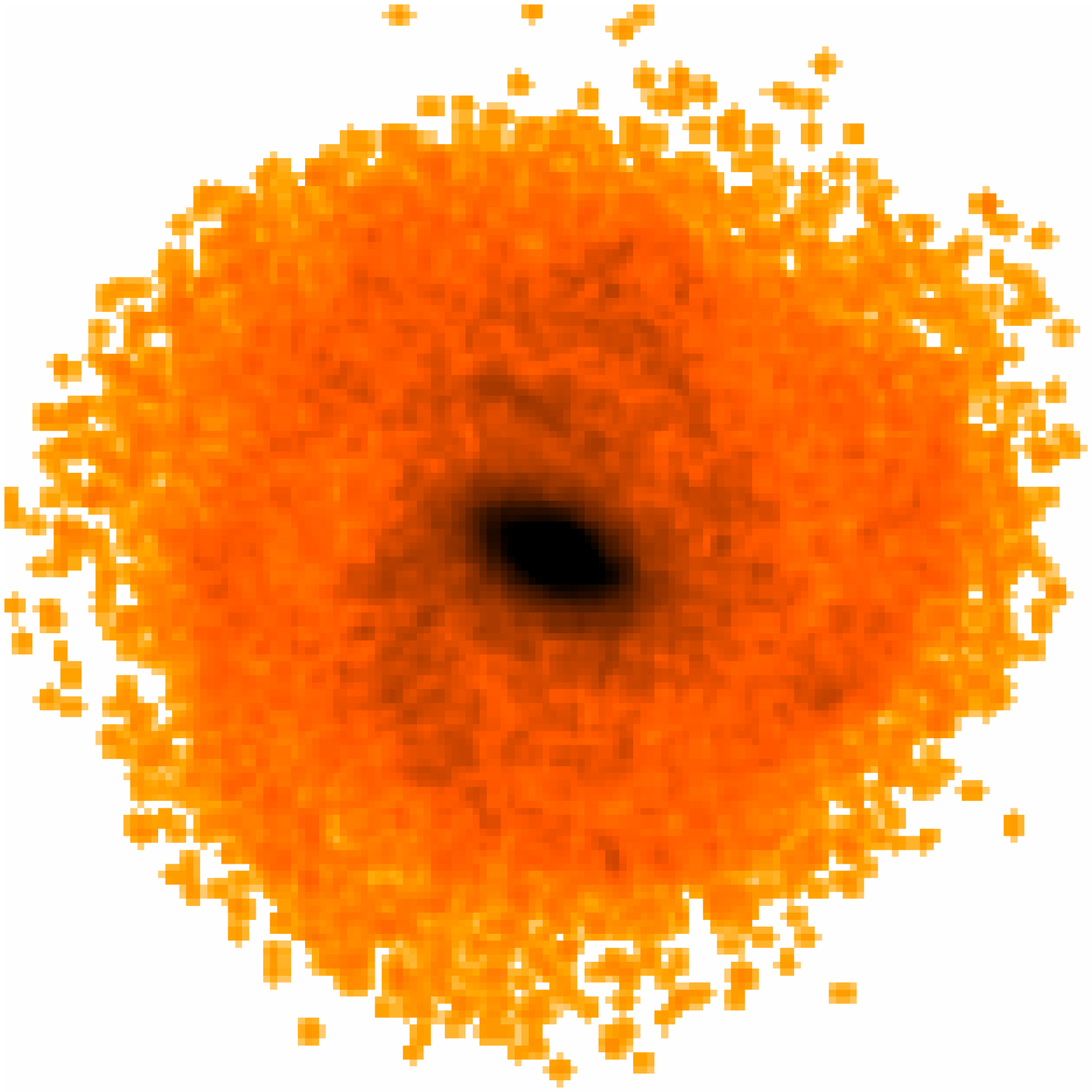}
\includegraphics[width=2cm,angle=0]{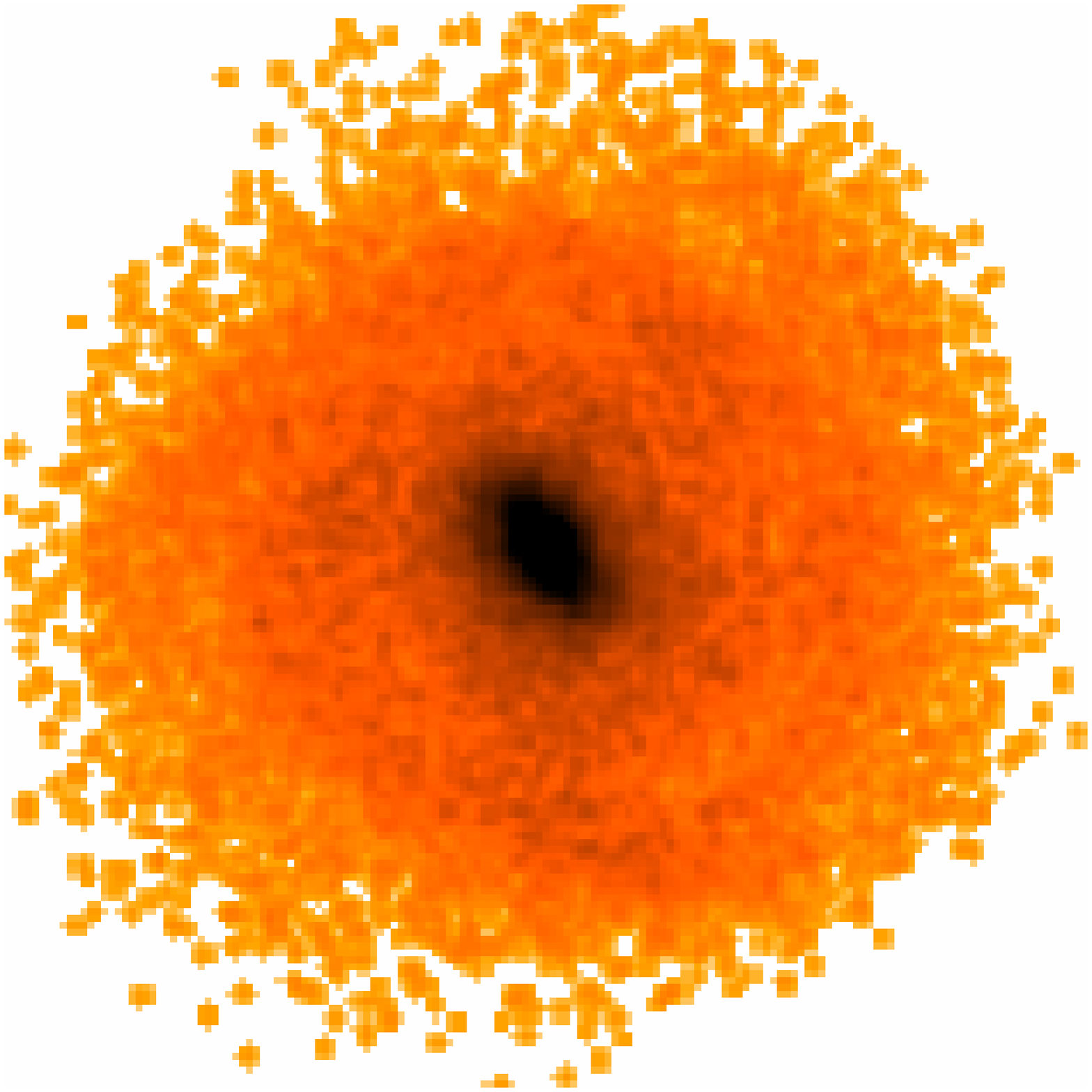}
\includegraphics[width=2cm,angle=0]{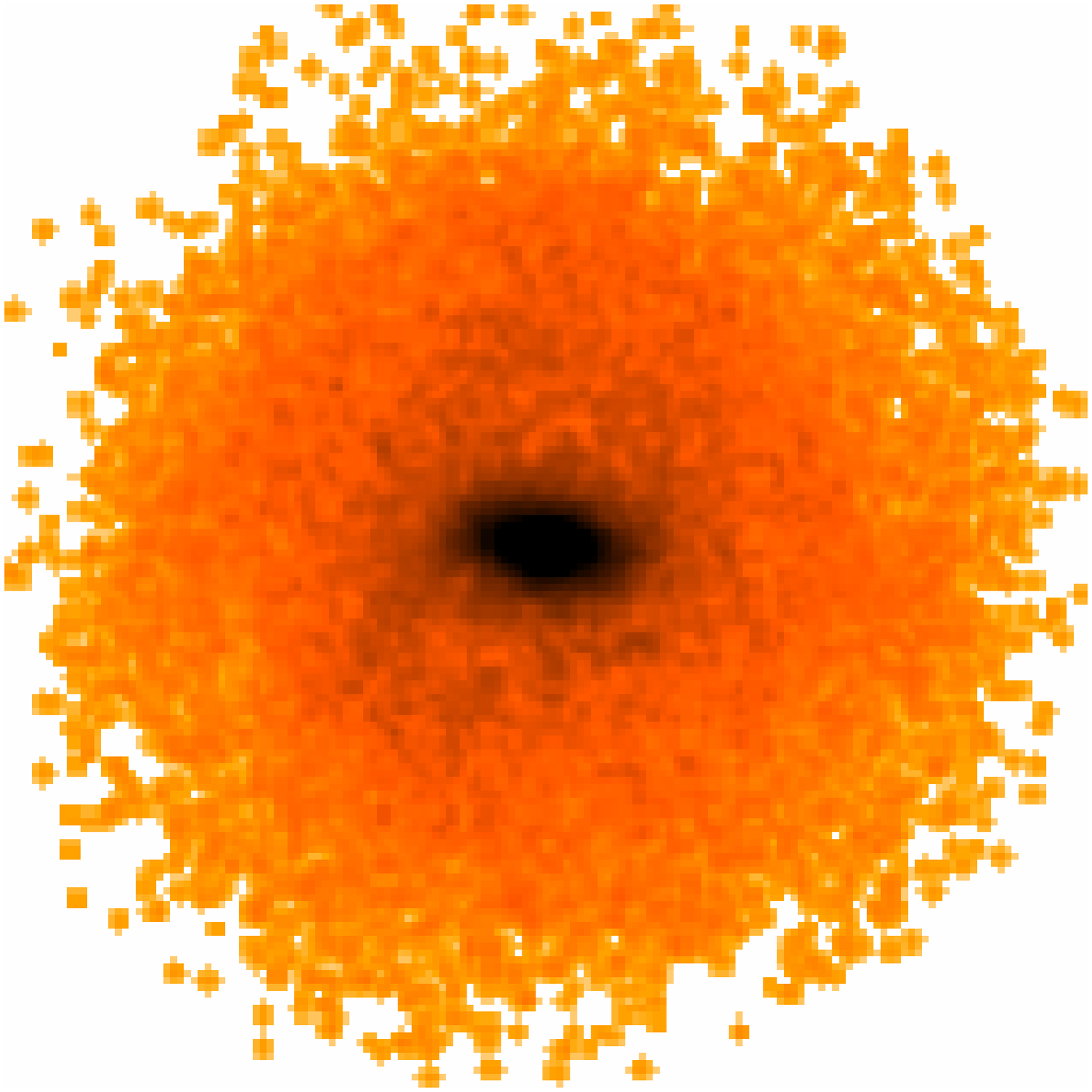}
\includegraphics[width=2cm,angle=0]{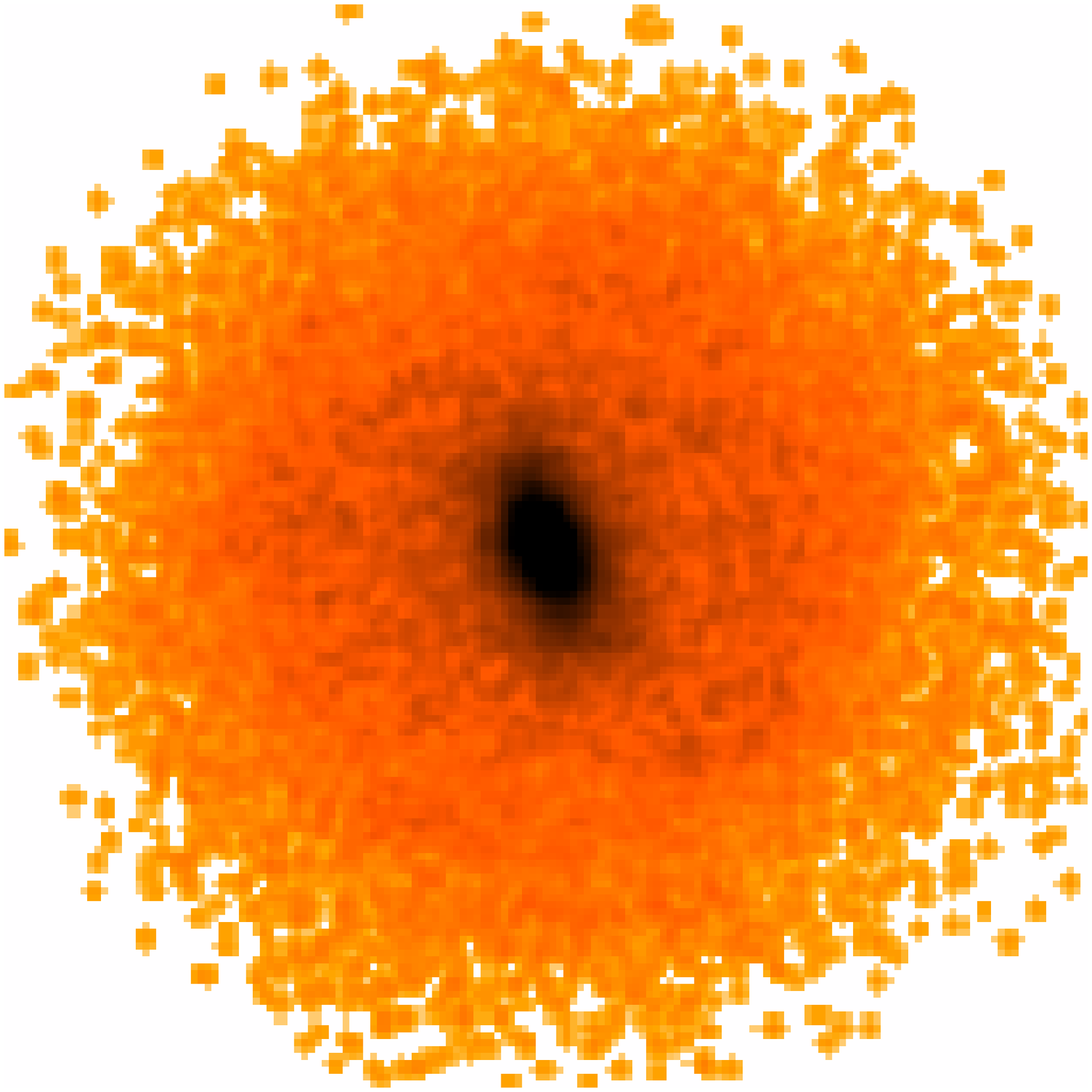}
\includegraphics[width=2cm,angle=0]{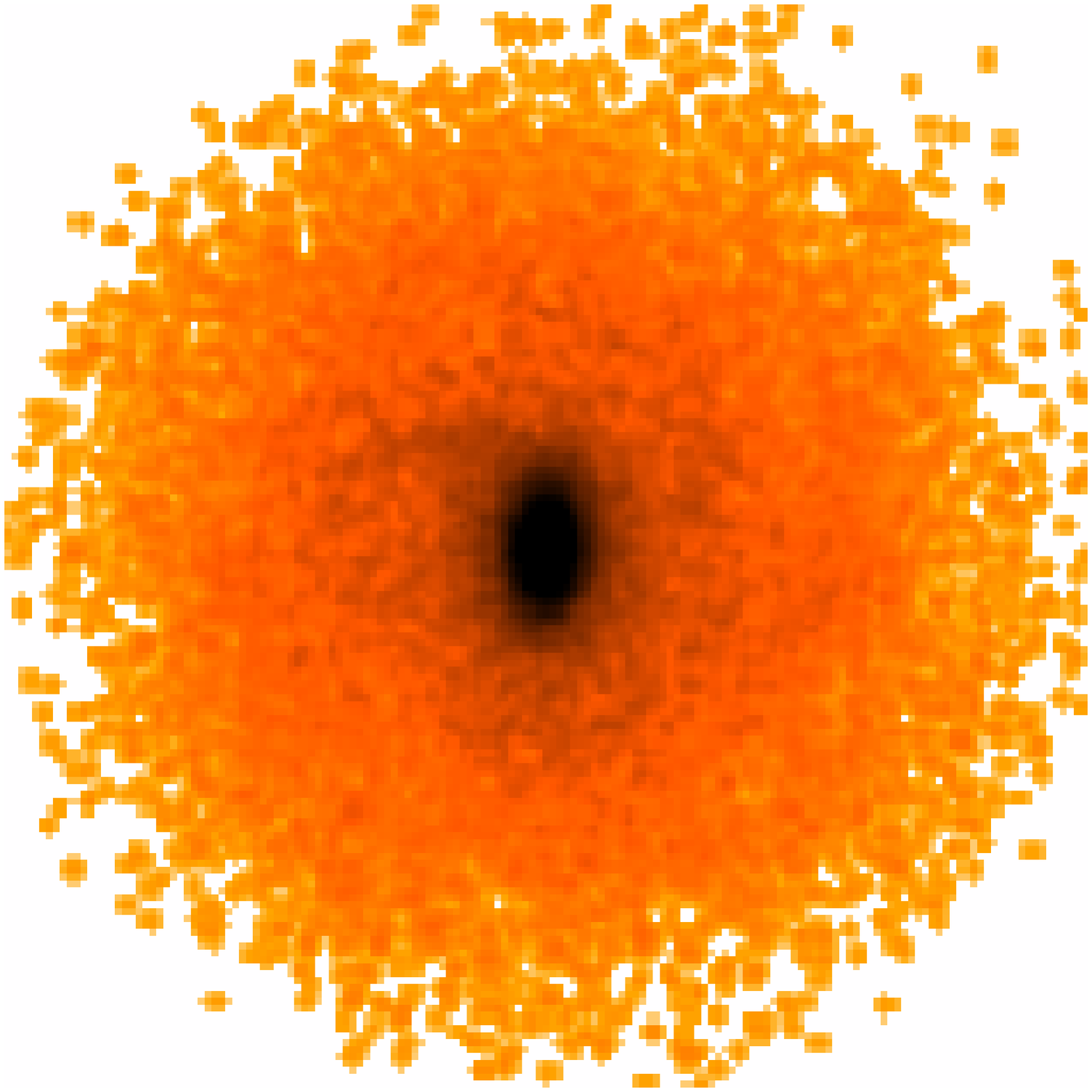}
\caption{From top-left to bottom-right: Maps of the stellar distribution of the gSb galaxy model evolved in isolation. Maps are shown for every 500 Myr over t=0 to t=3 Gyr. Each map is 40 kpc $\times$ 40 kpc in size.}
\label{fig:isostars}
\end{figure}

\section{Metallicity evolution in the isolated galaxy}\label{isolated}

Since the goal of this paper is to investigate the role interactions
and mergers play in the gas phase metallicity evolution, it is essential
to study the evolution of the gas properties of simulated galaxies that
did not experience a merger or interaction (i.e., an isolated galaxy).
This is essential to distinguish between secular processes that drive the
evolution of galaxies from the effects that occur during the interaction
and merger process.  For the isolated galaxy simulation, initial
metallicity profile (given in Eqn.\ref{gradient}) evolves rapidly in the
first 500 Myr of evolution, due to the compression of gas into density
waves and inflow into the central regions (Fig.\ref{fig:isogas}). At this
time, spiral arms and a bar are formed, as clearly shown in the maps of
the stellar distribution (Fig.\ref{fig:isostars}). The galaxy
models studied in this paper were not evolved in isolation before starting
the interaction. However, as shown in Figs. A1 and A2, tidally driven
gas inflows from the outer disk take place after the pericentre passage
(with our initial choice of the orbital parameters, this corresponds to
around 400 Myr from the beginning of the simulation) and in the merging
phase (between 1 and 3 Gyr with the exact timing depending on the orbital
characteristics). By this time -- 400 Myrs -- the metallicity profile
has already reached an equilibrium configuration (see Fig.A1), and the
galaxy has relaxed.

\end{appendix}

\end{document}